\documentclass[times,twocolumn,final]{elsarticle}

\usepackage{medima}
\usepackage{framed,multirow}

\usepackage{amssymb}
\usepackage{latexsym}

\usepackage{url}
\usepackage{xcolor}

\usepackage{hyperref}
\usepackage{subfigure}

\usepackage{amsmath}
\usepackage{soul}
\usepackage{threeparttable}
\usepackage{booktabs}

\definecolor{newcolor}{rgb}{.8,.349,.1}

\journal{Medical Image Analysis}

\begin{document}

\verso{Given-name Surname \textit{et~al.}}

\begin{frontmatter}

\title{Stain-Adaptive Self-Supervised Learning for Histopathology Image Analysis}%

\author[1,2]{Hai-Li \snm{Ye}\corref{cor1}}
\author[1,2]{Da-Han \snm{Wang}}
\cortext[cor1]{Department of Computer and Information Engineering, Xiamen University of Technology, Xiamen 361000, CN. E-mail addresses: yehl@mail.sustech.edu.cn (H. Ye).} 

\address[1]{Department of Computer and Information Engineering, Xiamen University of Technology, Xiamen 361000, CN}
\address[2]{Fujian Provincial Key Laboratory of Pattern Recognition and Image Understanding, Xiamen 361000, CN}

\received{1 May 2013}
\finalform{10 May 2013}
\accepted{13 May 2013}
\availableonline{15 May 2013}
\communicated{S. Sarkar}

\begin{abstract}
It is commonly recognized that color variations caused by differences in stains is a critical issue for histopathology image analysis. Existing methods adopt color matching, stain separation, stain transfer or the combination of them to alleviate the stain variation problem. In this paper, we propose a novel Stain-Adaptive Self-Supervised Learning(SASSL) method for histopathology image analysis. Our SASSL integrates a domain-adversarial training module into the SSL framework to learn distinctive features that are robust to both various transformations and stain variations. The proposed SASSL is regarded as a general method for domain-invariant feature extraction which can be flexibly combined with arbitrary downstream histopathology image analysis modules (e.g. nuclei/tissue segmentation) by fine-tuning the features for specific downstream tasks. We conducted experiments on publicly available pathological image analysis datasets including the PANDA, BreastPathQ, and CAMELYON16 datasets, achieving the state-of-the-art performance. Experimental results demonstrate that the proposed method can robustly improve the feature extraction ability of the model, and achieve stable performance improvement in downstream tasks.
\end{abstract}


\end{frontmatter}



\section{Introducing}

Pathological diagnosis(\cite{Bjartell2005}; \cite{EmadMaysa2008}; \cite{SymmansPeintinger2007}) plays a critical role in clinical medicine because it provides objective evidence for the diagnosis, classification, and treatment of diseases, as well as for the judgment of disease progression, prognosis, and efficacy. Pathology is the process and principle of the occurrence and development of a disease, including the causes and regulations of the event of a condition, as well as the changes in the structure, function, and metabolism of cells, tissues, and organs during the process of an illness and their rules. In addition to providing diagnostic information, the phenotypic information contained in histology slides can be used for prognosis. Features such as nuclear atypia, degree of gland formation, presence of mitosis, and inflammation can all be indicative(\cite{EmadMaysa2008}) of how aggressive a tumor is and may also allow predictions to be made about the likelihood of recurrence after surgery.

The analysis of whole-slide digital pathology images (WSIs) (\cite{GurcanBoucheron2009}) is a challenging task because complex background with noises, the nuclei overlapping, variations in staining, etc. Among these problems, the problem of stain variation has been generally recognized as a critical issue that may deteriorate the performance of the system(\cite{RabinovichAgarwal2003}; \cite{BejnordiLitjens2016}; \cite{BenTaiebHamarneh2018};), and has attracted much attention in recent years. The resolution of WSIs exceeds tens of thousands of pixels, so it is necessary to use a sliding window to divide them into patches during analysis. The pathological features of microscopes at different magnifications may be all-important for a task. Therefore information from multiple scales needs to be integrated. When the proportion of lesion area in WSIS is tiny, the lesion information reflected by the pixels of the relevant area is minimal.All in all, the Analysis of WSIs is challenging in the following two respects. The first difficulty is that staining conditions vary greatly depending on the specimen and the hospital from which the sample was taken. Therefore, pathologists perform tumor region identification and subtype classification by carefully considering the different staining conditions. The last difficulty is that Pathological image data is relatively scarce. This is mainly due to the scarcity of some problematic cases and the high maintenance cost of WSIs. In addition, the current pathological image data sets are relatively independent, and data collection and labeling are carried out for different organs and lesions. Sample imbalance is more severe in pathological image analysis compared with other medical image analysis fields. However, some current work based on transfer learning and stain normalization successfully extract WSIs invariant features. But these methods are more complex and not applicable in the process of multiple transfers.

To address these common problems, we propose a simple and effective solution called Stain-Adaptive Self-Supervised Learning (SASSL) for Histopathology Image Analysis. Our goal is to offer a novel self-supervised method to achieve model adaptation to staining while extracting potential invariance features of WSIs. It reduces the impact of pathological image staining differences and can also use additional tasks to mine its monitoring information from large-scale pathological image data of different stains. The model aligns the distribution of all staining stains in common feature space to extract the invariant representation of the common staining stains. In addition, this module mines the supervision information of pathological image data itself, and the consistent features in pathological images can be learned like the self-supervision process. Its function is to extract the invariant features in the pathological image while aligning the staining stains of the pathological image to improve further the performance of the downstream model of specific pathological image analysis. Because each downstream is considered to have a particular distribution of features, we do not directly replace the downstream feature extraction module as in the general self-supervision process. Instead, we want the generic feature extractor as an auxiliary branch. Its function is to extract the aligned invariant features from pathological images, and these features can further enrich the feature extraction ability of downstream models.

\begin{figure*}[ht]
	\centering
	\includegraphics[width=\linewidth,scale=1.00]{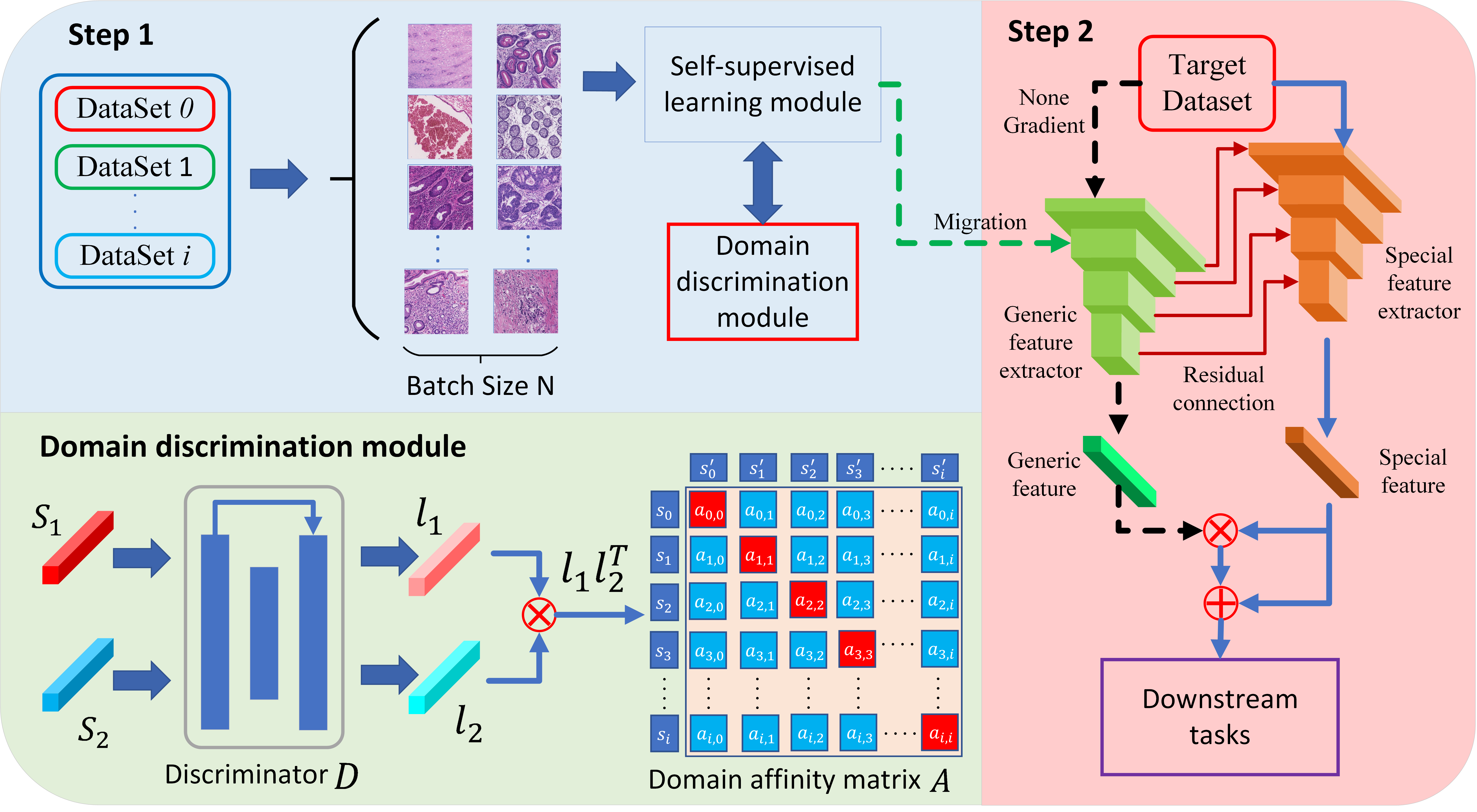}
	\caption{stain-adaptive self-supervised learning (SASSL) architecture (Please note that the input image size does not need to be fixed value). In step 1 (blue area), samples from multiple pathological image data sets are input into SASSL for stains adversarial self-supervised learning. Compared with traditional self-supervised methods, we introduce a stains discrimination module (green area). For a detailed description of the SASSL process, please refer to Chapter 3.2. In step 2, the weights learned by SASSL will be transferred to downstream tasks to improve the feature extraction ability of the model. For detailed description, please refer to Chapter 3.3}
	\label{Fig.1}
\end{figure*}

\section{Related Work}
\label{Chapter 2}

In our review of related works, we first introduce three major canonical deep learning models based on the nature of tasks that are solved in digital histopathology: classification, regression, and segmentation-based models, and then we summarize the self-supervised learning based on contrastive learning and pathological image stain normalization. Finally, we will explain the difference and advancement of our work.

\textit{Pathological Image Classification:} Image classification is a classic computer vision problem, Compared with the general visual image, the contour, and texture contained in the image block of WSIS are more complex. As the number of layers of the neural network increases, the problem of gradient vanishing and gradient explosion will be more obvious. In \cite{HeZhang2015}, the authors propose a residual learning method, which uses the deep layer as the shallow layer Identity mapping to prevent the gradient from disappearing or exploding. In addition to similar structures(\cite{HuangLiu2017}; \cite{TanLe2019};\cite{DingZhang2021};), others have a series of models based on attention mechanism include (\cite{HuShen2018}; \cite{WooPark2018}; \cite{GaoCheng2021}; \cite{HouZhou2021}). However, the global classification of pathological images is more complicated than the scene image classification. The commonly used pathological image classification (\cite{CampanellaHanna2019}; \cite{GaoWang2017}; \cite{TellezBalkenhol2018};  \cite{XuLiu2019};) is based on multiple instances of a learning-based deep learning system that uses only the reported diagnoses as labels for training, thereby avoiding expensive and time-consuming pixel-wise manual annotations.

\textit{Pathological Image Regression:} This kind of research focuses on the detection or localization of objects by directly regressing the likelihood of a pixel being the center of an object (e.g., cell or nucleus center). Scene image regression first assumes a linear relationship between the pixel values of two images in different periods. The regression and classification models are very similar in structure; the major difference lies at the end of the model. At present, the main research is in the target detection how to more accurately return to the area of the object (\cite{XieTu2017}; \cite{NaylorMarick2019}; \cite{ChenWang2016}; \cite{KashifRaza2016};). Detection of cells or nuclei in histopathology images is challenging due to their highly irregular appearance and their tendency to occur as overlapping clumps, which results in difficulty in separating them as a single cell or a nucleus (\cite{NaylorLae2019}; \cite{XieKong2015}; \cite{GrahamChen2019}). Therefore, end-to-end regression models can be used directly for tumor load assessment(\cite{GeertPeter2018}; \cite{PeterOscar2019}; \cite{BabakMitko2017};) and lymphocyte assessment (\cite{MohammadPeikari2017}; \cite{TahsinSharma2019}; \cite{LomacenkovaArandjelovic2021}).

\textit{Pathological Image Segmentation:} In recent years, the development of deep neural networks has achieved many breakthroughs in automatic histopathology image segmentation. FCN(\cite{LongShelhamer2015}) promotes the use of end-to-end convolutional neural networks in semantic segmentation problems. U-Net(\cite{RonnebergerFischer2015}) model constructs a complete set of encoder-decoder, which is widely used in medical image segmentation. And other follow-up work (\cite{AbdulkadirLienkamp2016}; \cite{MilletariNavab2016}; \cite{ZhouSiddiquee2018}; \cite{ChenZhu2018}). And also work on the features of the lesions, such as the liver cancer(\cite{KimJang2021}; \cite{SchmitzMadesta2021};),Breast cancer(\cite{LiuXu2019}; \cite{KianiUyumazturk2020}), lung cancer (\cite{NicolasSantiago2018}; \cite{WangChen2020};).In addition to segmentation based on global size, there are also segmentation based on nuclear level (\cite{SongTan2019}; \cite{HuTang2019};). Many researches also aim at optimizing loss to improve segmentation quality (\cite{MilletariNavab2016}; \cite{KervadecBouchtiba2021};). Most deep learning methods in digital pathology are applied on small-sized image patches rather than the entire WSI, restricting the model’s prediction ability to a narrow field of view.

\textit{Self-Supervised Contrastive Learning:} In recent years, self-supervised learning(\cite{JingTian2019}; \cite{LiuZhang2020}; \cite{ChenHe2021}; \cite{ChenFan2020}) as a new representational learning method has achieved many achievements. Self-supervised learning mainly uses auxiliary tasks to mine its supervised information from large-scale unsupervised data. Self-supervised trains the network with the constructed supervisory information so that the valuable representations for downstream tasks can be learned. After self-supervised training, the learned visual features can be further migrated to downstream tasks as pre-training models to improve performance and overcome over-fitting. We summarize them into three main categories according to their objectives: generative(\cite{WangHuang2014}; \cite{FabiusAmersfoort2015}; \cite{XuGopale2020};), contrastive(\cite{ChenKornblith2020v1}; \cite{ChenKornblith2020v2}; \cite{HeFan2020}; \cite{ChenFan2020}; \cite{GrillStrub2020}), and adversarial(\cite{RadfordMetz2016}; \cite{ZhangIsola2016}; \cite{ZhangIsola2017};). The generation-based or adversarial-based approach focuses more on the details of the pixel than on the more abstract underlying factors. Therefore, the method based on a contrastive pays more attention to the global semantic features of images and can learn more complex potential representations of images.

\textit{Pathological Image Stain Normalization:} Stain normalization techniques have been an important preprocessing before most computer-aided diagnostic tasks, as slides from different institutions and even slides within the same institution but different batches may vary drastically in stain style. Many successful stain normalization techniques have been proposed in the discipline of computational pathology. The initial approach attempts to subtract each color channel's mean and then divide by the standard deviation\cite{Harshal}. Conventional algorithms have demonstrated their effectiveness in the past, like a non-linear mapping approach that employs image-specific color deconvolution\cite{Macenko}, or an estimation of the stain vectors using singular value decomposition (SVD) geodesic-based stain normalization technique\cite{Reinhard}, and many others\cite{Vahadane, Saad}. However, due to the reliance on an expertly selected target image, the conventional methods are no more applicable to convolutional neural networks. More recently, generative adversarial networks have been widely adopted to reduce color variations between multiple data centers\cite{KeJing}. Specifically, the StainGAN\cite{Staingan} have demonstrated their outstanding experimental results concerning stain separation and image information preservation.

In this paper, we propose a simple and effective solution called Stain-Adaptive Self-Supervised Learning (SASSL) for Histopathology Image Analysis. Our goal is to propose a novel self-supervised method to achieve model adaptation to staining while extracting potential invariance features of WSIs. The key contributions of our paper are:
\\
\\

\begin{itemize}
	\item The proposed SASSL method integrates a stains discriminator module into the SSL framework to perform adversarial learning. This method improves the adaptive ability of the model to WSIs staining changes and enables the model to learn to extract potential invariance features in WSIs.
	\item Our SASSL method can be regarded as a general model for invariant feature extraction which can be flexibly combined with arbitrary downstream histopathology image analysis modules (e.g. Classification/ Regression/Segmentation).
\end{itemize}

\begin{figure*}[ht]
	\centering
	\subfigure[MoCoV2 structure overview diagram, MoCov2 is a representative of the self-supervised learning framework structure based on positive and negative samples.]{
		\includegraphics[width=0.45\linewidth]{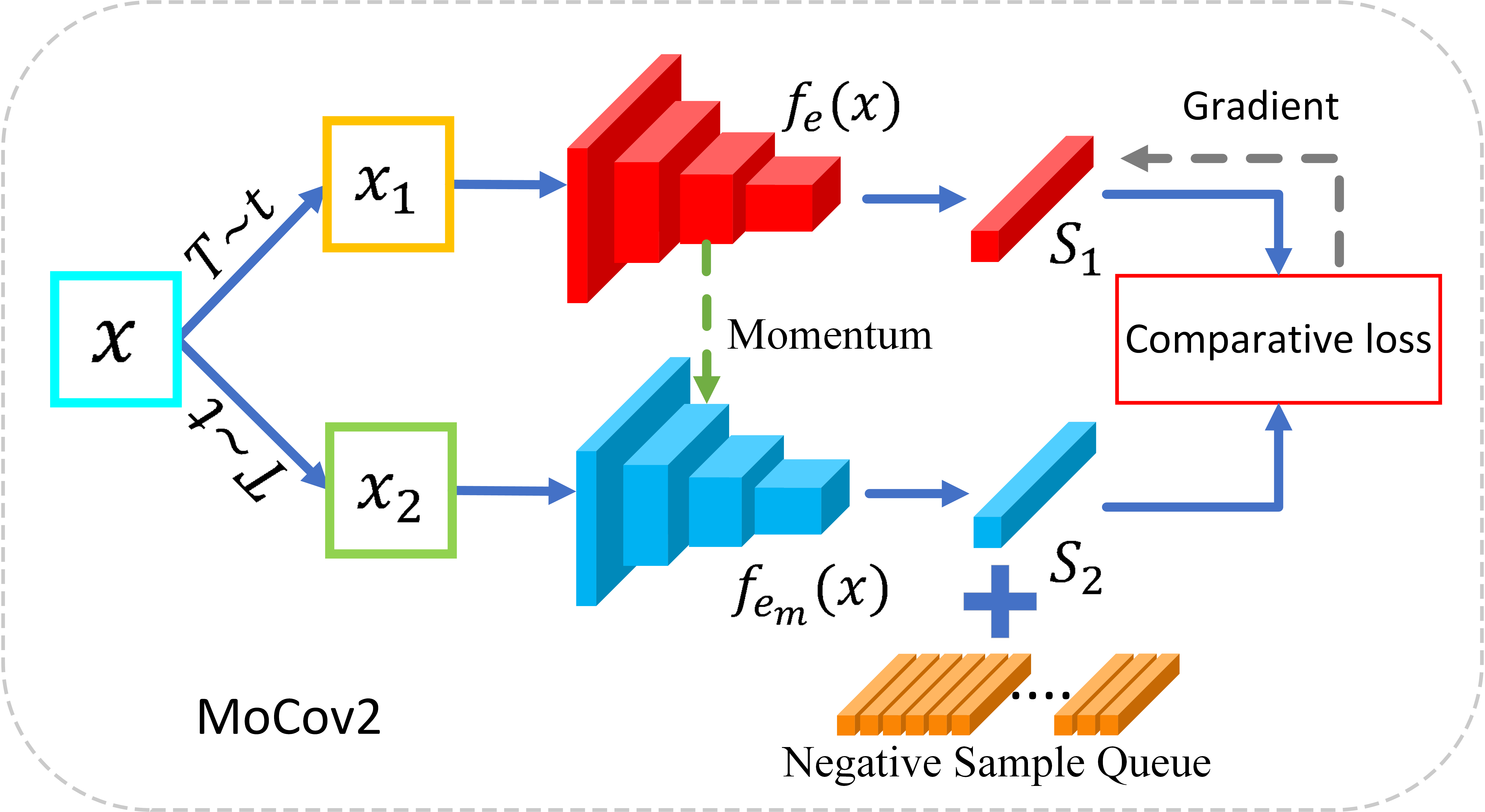}
	}
	\centering
	\subfigure[SimSiam structure overview diagram, SimSiam is a representative of the self-supervised learning framework structure based on only positive samples.]{
		\includegraphics[width=0.45\linewidth]{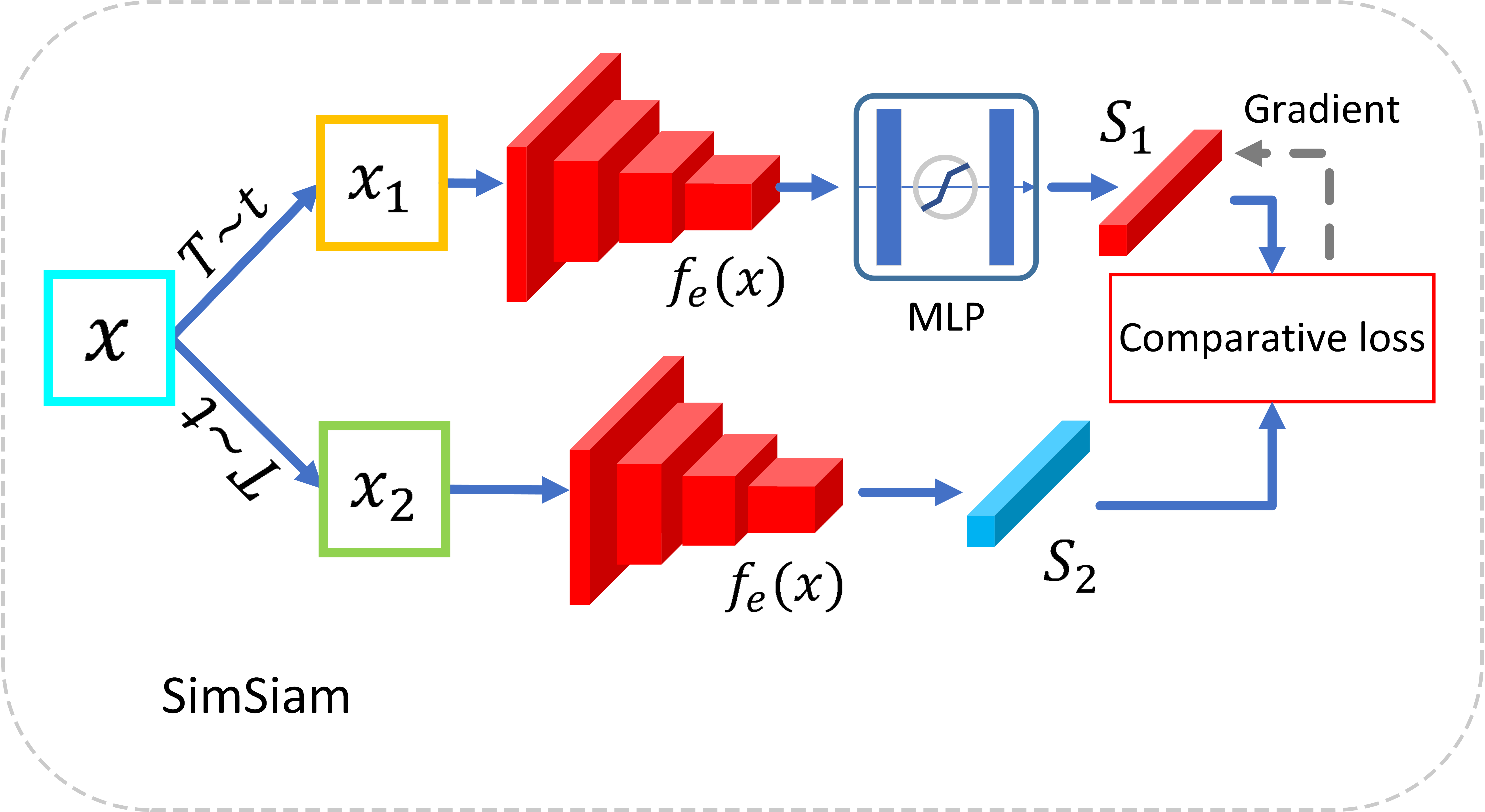}
	}
	\caption{MoCov2 and SimSiam structure overview diagram.}
	\label{Fig.2}
\end{figure*}

\section{Methodology}
Fig.1 illustrates the overview of the proposed Stain-Adaptive Self-Supervised Learning(SASSL) framework for the Pathological image analysis. It mainly consists of two steps: (1) In the first stage, we used multiple pathological image data sets for self-supervised learning without using downstream task supervised labels. We introduce a stains discriminant module so that the self-supervised module realizes the stains alignment of pathological images in the high-dimensional feature space and mines the potential standard features in pathological images. (2) We treat the encoder learned in the first stage as a universal feature extractor in the second stage. The generic feature extractor can extract general features in pathological images and parallel to the special feature extractor of the original downstream framework to extract and fuse image features. This process can be embedded into any downstream analysis framework of pathological images, which improves the analysis capability of the model robustly and has good mobility.

\subsection{Self-Supervised Learning Framework}
\label{Chapter 3.1}
In order to realize stain-adaptive self-supervised learning of general data of pathological images, we first need to integrate pathological image data sets from different task scenes. We selected the three classic task scenes in the visual analysis of pathological images: classification, regression, and segmentation. The core of stains adaptation is to make the model unaffected by the stains changes of WSIs and achieve robust feature extraction for any stains situation. This idea is similar to the fine-grained identification scene, which contradicts small inter-class and large intra-class distances. We treat multiple pathological image datasets as a whole, including $N$ WSI slices. We denote the multi-source dataset as $\{(X_{n},Y_{n})\}_{n=1}^{N}$, where $X_{n}$ is WSI and $Y_{n}$ is downstream task label. The self-supervised learning in the first phase of our framework is an unsupervised process that does not need to use labels for downstream tasks. When the downstream task is a classification, $Y_{n}$ is the single hot coding of the classification label. Similarly, $Y_{n}$ corresponds to linear predictive value and segmentation mask in regression and segmentation tasks, respectively. Since each WSI is too huge to feed into a CNN directly, a patch-based approach is usually employed. In addition, different downstream tasks have different requirements for the input patches size, so we hope the model can accept multi-scale patches. In this paper, we consider patches with $512\times512$ pixels. Therefore, each WSI can be regarded as a stain type, and the patches generated are the sample of this stain. Patches is denoted by $P_{n}$ for $ {n} \in [N]$ and each patch $ {p} \in P_{n}$ Belongs to the corresponding WSI. stain-adaptive self-supervised learning aims to align the features of patches in different stains in high-dimensional space and mine the potential common features in pathological images.

The structural paradigm of SASSL is modeled on self-supervised learning based on contrastive learning. So, in the next section, we will briefly summarize the self-supervised contrast learning paradigm in preparation for the introduction of our stains antagonism module. Our approach is an extension of the underlying paradigm of comparative self-supervised learning that applies to approaches of the same paradigm type. Several SSL methods can be adopted as the base framework. Contrastive learning aims to "learn to compare" through a Noise Contrastive Estimation (NCE) objective formatted as:
\begin{equation}
			L={\mathbb{E}}_{{x},{x}^+,{x}^-}[-log(\frac{{e}^{f(x)^Tf(x^+)}}
			{{e}^{f(x)^Tf(x^+)}+{e}^{f(x)^Tf(x^-)}})],
\end{equation}
where ${x}^+$ is positive sample to ${x}$, ${x}^-$ is negative sample to ${x}$ and $f$ is an encoder (in this paper, it is a general feature extractor). The similarity measure and encoder may vary from task to task, but the framework remains the same. In general, with more negative sample pairs involved, we have the InfoNCE formulated as:
\begin{equation}
	L={\mathbb{E}}_{{x},{x}^+,{x}^-}[-log(\frac{{e}^{f(x)^Tf(x^+)}}
	{{e}^{f(x)^Tf(x^+)}+\sum\nolimits_{k=1}^K{e}^{f(x)^Tf(x^-)}})],
\end{equation}

The deepening of related research mainly focuses on two process paradigms of contrastive learning. The first paradigm is mainly to increase the proportion of negative samples to strengthen comparative learning. We take MoCov2(\cite{ChenFan2020})(Fig.2a) as an example. In MoCov2, researchers further develop the idea of leveraging instance discrimination via momentum contrast, substantially increasing the negative samples' amount. The data goes through two or more different enhancements to get different views. These views have differences in shallow features but are consistent in deep representation. Where $f_e(x)$ uses gradient descent training, $f_{e_m}(x)$ updates momentum according to $f_e(x)$, maintaining $f_e(x)$ and $f_{e_m}(x)$ similar but different state. After the view of the same sample, $X$ is encoded by the encoder to get $S_1$ and $S_2$; the same sample vector features should be aligned. After each iteration, the previous sample is entered into the negative sample queue. Finally, there are positive sample pairs between $S_1$ and $S_2$, and negative sample pairs between $S_1$ and negative sample queue, using the Equation 2 InfoNCE loss to update the model parameters. The second paradigm discarded the form of negative sampling and achieved better results in self-supervised learning than InfoNCE(\cite{Gutmannrinen2012}). We take SimSiam(\cite{ChenHe2021})(Fig.2b) as an example. Simsiam abandons momentum update and changes to the twin network structure. An additional set of MLP is added to the $f_{e}(x)$ encoder branch and turns off the gradient of another branch. Similarly, different views are passed through two encoders to get $S_1$ and $S_2$. Simsiam followed the regression paradigm to design the loss function:

\begin{equation}
	L=1-\frac{<S_1, S_2>}{{||S_1||}_2\cdot{||S_2||_2}},
\end{equation}

\subsection{Stains Discriminator Module}
Both paradigms finally generate eigenvectors $S_1$ and $S_2$ of different views of samples. For pathological images, the previous comparative learning process can only learn the potential characterization of patches but cannot realize the alignment of the staining stains between patches in high-dimensional space. For pathological images, the previous comparative learning process can only learn the potential characterization of patches but cannot realize the alignment of the staining stainss= between patches in high-dimensional space. So we designed an A stains discriminator module (Fig.1, the red box) and integrated it into the SSL framework. Details of the stains discriminator module are shown in Fig. 1. Light green area. First, the outputs of the SSL by $S_1$ and $S_2$ are input to a Discriminator $D$ that consists of linear layers and coupled LeakyReLu layers, as shown in the figure. In particular, a shortcut connection is added after the last LeakyReLu for residual learning, effectively avoiding the vanishing gradient problem. $l_1$ and $l_2$ denote the feature vectors extracted from $S_1$ and $S_2$ that are from different stains augmentations, respectively.

\begin{figure}[ht]
	\centering
	\subfigure[No same source patches]{
		\includegraphics[width=0.45\linewidth]{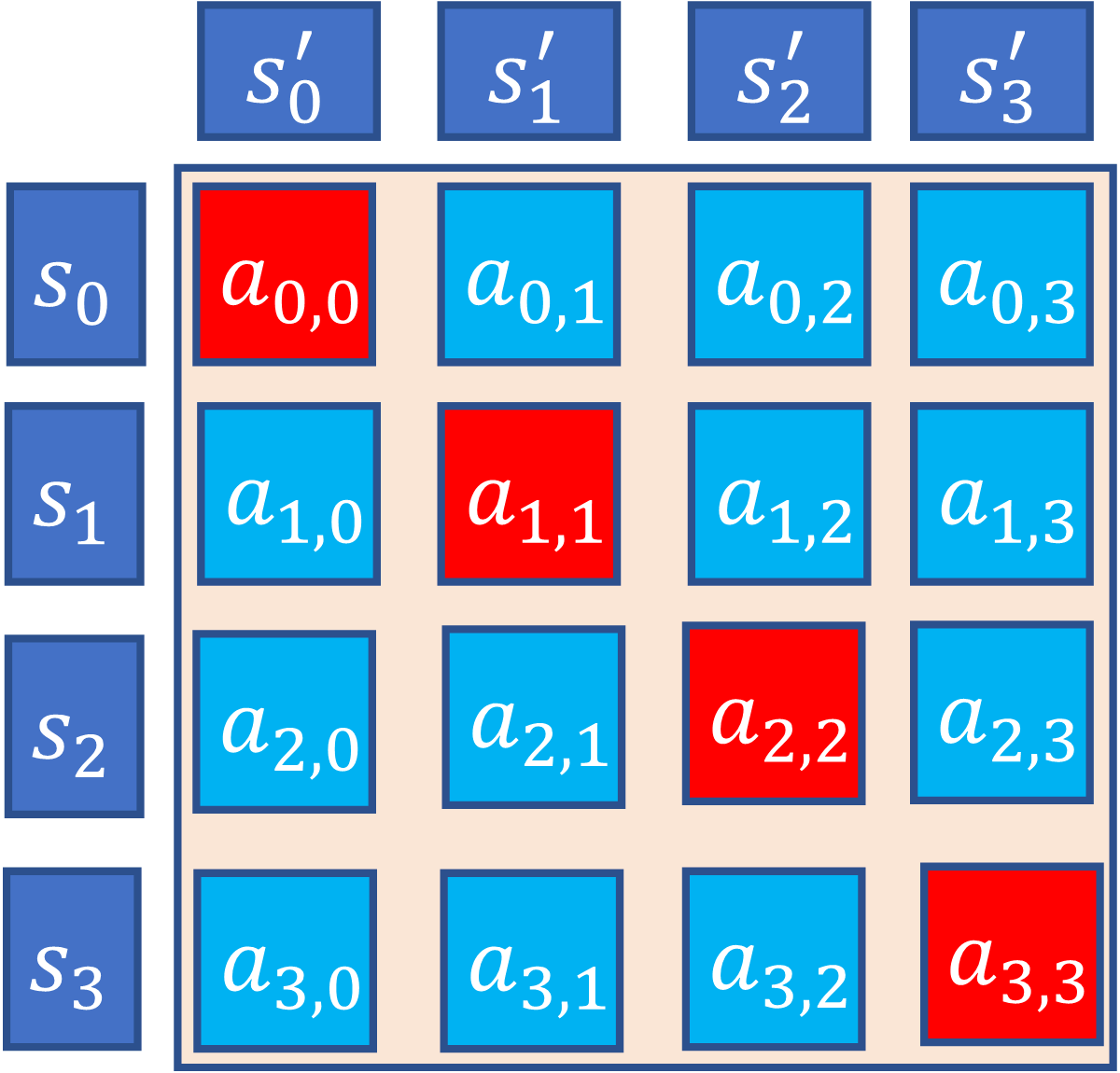}
	}
	\centering
	\subfigure[Two patches are the same source]{
		\includegraphics[width=0.45\linewidth]{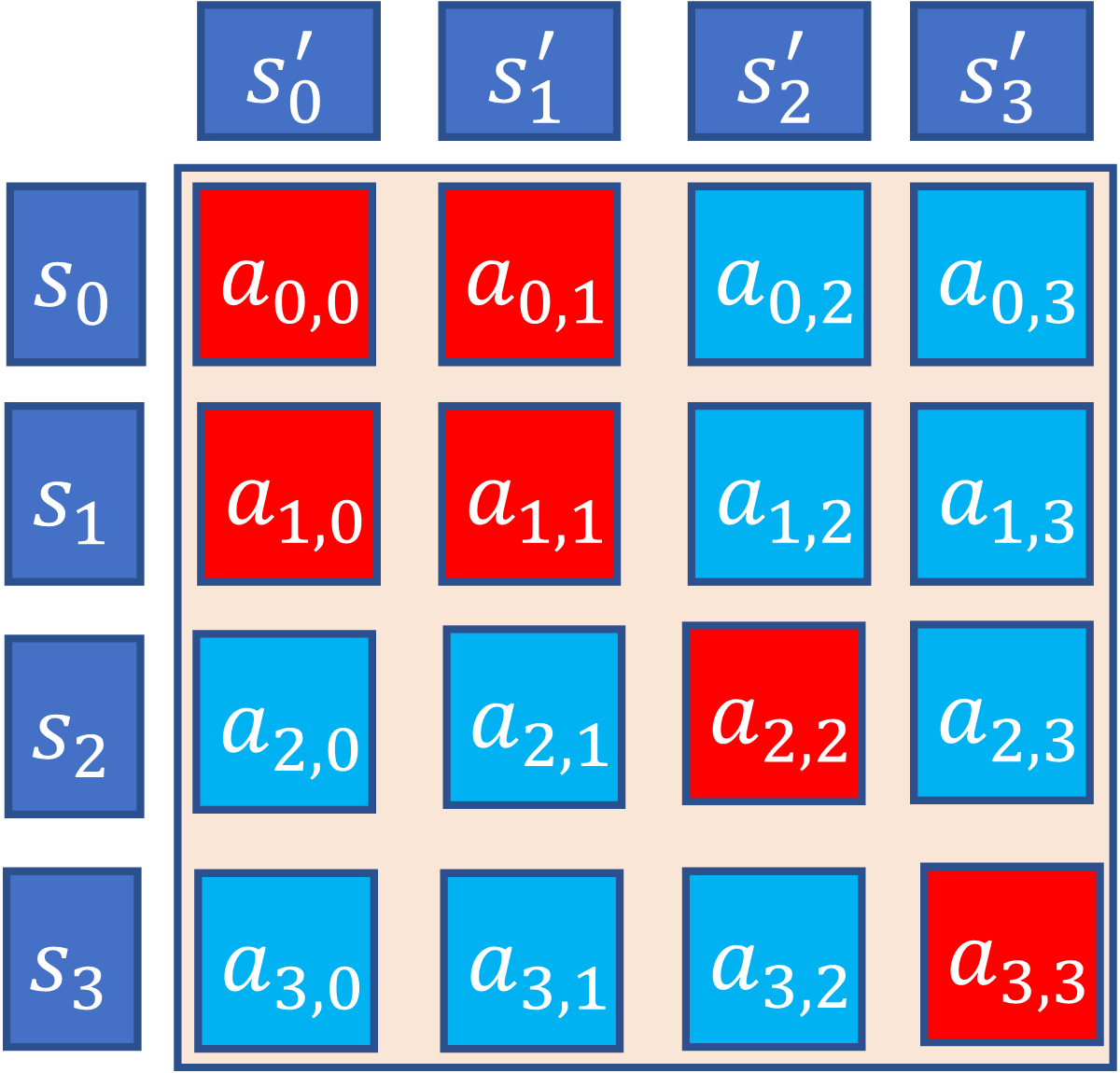}
	}
	\centering
	\subfigure[Thre patches are the same source]{
		\includegraphics[width=0.45\linewidth]{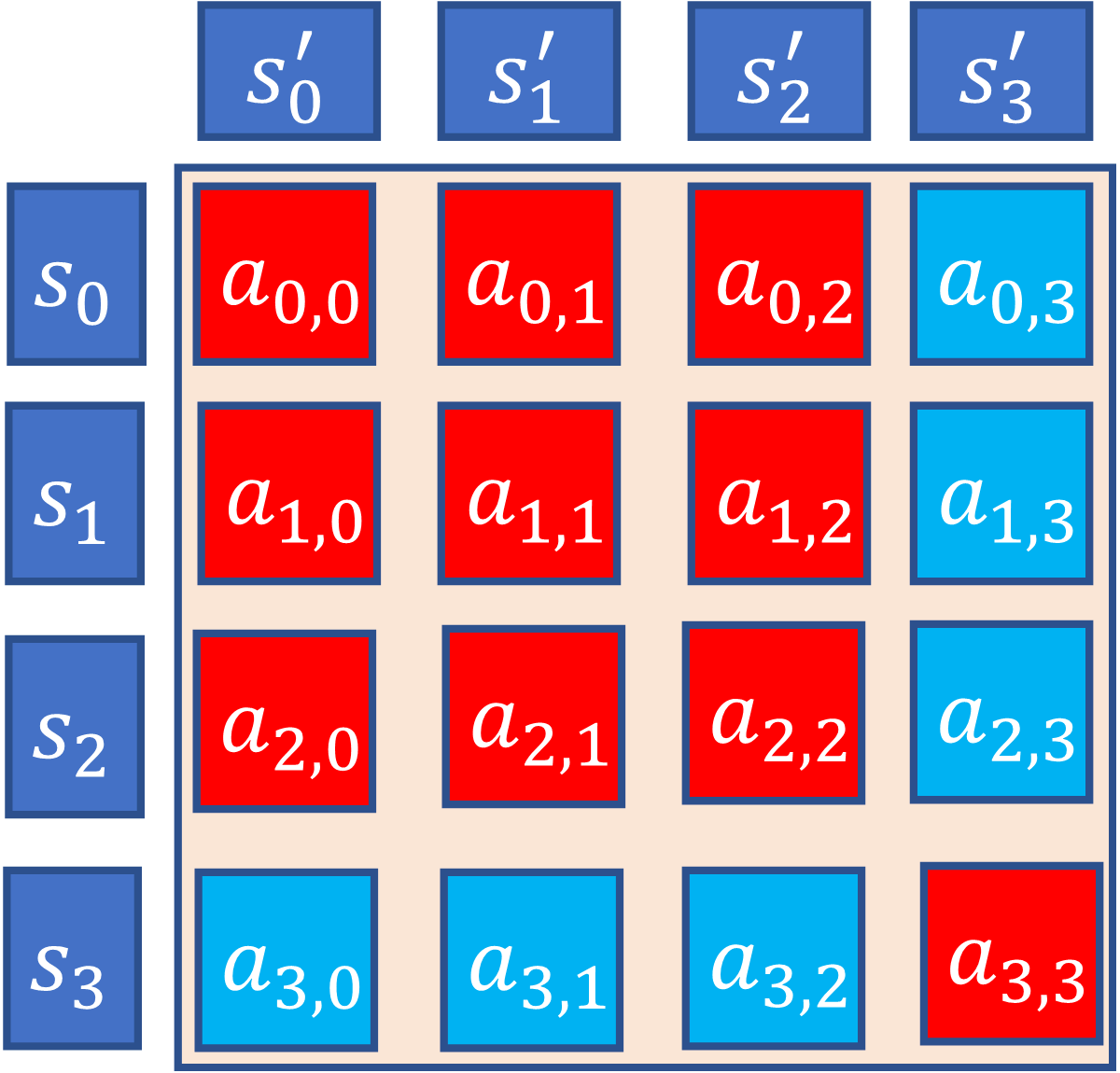}
	}
	\centering
	\subfigure[All patches are the same source]{
		\includegraphics[width=0.45\linewidth]{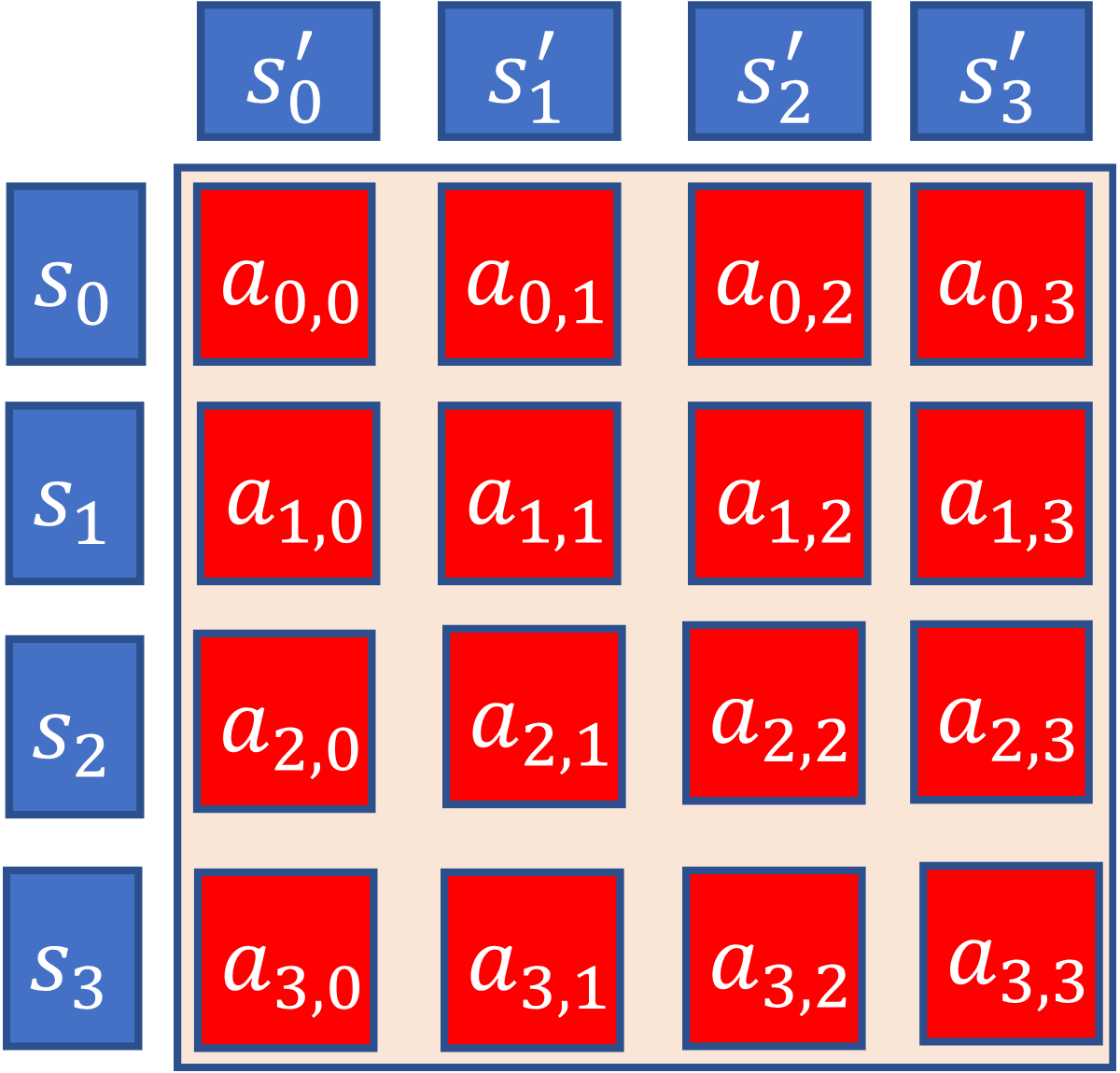}
	}
	\caption{The stains affinity relationship. We take the example of Mini-batch with N as 4 to show the affinity relationship between patches under different conditions.}
	\label{Fig.3}
\end{figure}

\textit{Stains Affinity Matrix}. We propose a stains affinity matrix for one mini-batch with $N$ images to compute the affinity among different patch stains. The stains affinity matrix $A:{\mathbb{R}^{N,N}}$ is obtained by multiplying ${l_1}:{\mathbb{R}^{N,d}}$ and ${l_2}^T:{\mathbb{R}^{d,N}}$, where ${a_{i,j}}$ represents the stains affinity of the $i$ and $j$ patch. For the stain discriminator, it is desired that, for the same image with a different augmentation view, the affinity between them is high. On the contrary, for different images, the affinity is low. When all patches in mini-batch are from different stains, the diagonal of the affinity matrix (the red block) is desired to be high while the others (the blue block) are desired to be low. In order to better illustrate this concept, we take the mini-batch $N$ as four as an example, as shown in Fig.4. Only the blocks in the diagonal region are affinities when there are no patches of the same source. When the number of homologous patches increases, the affinity blocks increase. The relational matrix $D$ is written as: $R=({r}_{i,j})=\begin{cases}0, p_i \neq p_j \\ 1, p_i = p_j \end {cases} $.

\textit{Stain-Adaptive Training}. To reduce the model's sensitivity to the stains, the adversarial training was carried out for the affinity matrix. On the contrary to the stains discriminator, for the generator, i.e., the feature learning of SSL, it is desired that all the images have high affinities despite what the augmentation/stains are. We treat the SSL part as generator $G$ and the stains counter module as $D$. Using the idea of GAN training, the Stain-Adaptive training object function will be:
\begin{equation}
	\min\limits_{\theta_G, \theta_D},\max\limits_{\theta_D} {L}_{sal}(G,D) + {L}_{ssl}(G),
\end{equation}
Further, we specifically write the loss expression of the antagonistic training process:
\begin{equation}
	L_{G} = L_{ssl} - \frac{||(1-R)A||_1}{||(1-R)||_1} -\frac{||RA||_1}{||R||_1} = L_{ssl} - \frac{||A||_1}{N^2},
\end{equation}
\begin{equation}
	L_{D} = \frac{||(1-R)A||_1}{||(1-R)||_1} -\frac{||RA||_1}{||R||_1},
\end{equation}
$L_{ssl}$ in the above formula can be replaced by the self-supervised loss we introduced in Chapter 3.2. The above GAN training objective is regarded as a saddle point optimization problem, and gradient-based methods often accomplish the training. $G$ and $D$ have trained alternately from scratch so that they may evolve together. Where $A$ is the affinity matrix of the model output, and R is the relationship matrix. $R$ represents patches of red homology in Fig. 4, and $1-R$ means patches of blue from different sources. For Eq.5, the goal of the generator $G$ is to ensure that the patch feature vectors of different colors are aligned simultaneously as self-learning. Therefore, the affinity of the feature vectors for any patch should be close (red and blue). Then all the values in the affinity matrix $A$ should be as large as possible so that the discriminator $G$ cannot discriminate the affinity of different slices to achieve the purpose of adaptation, that is, $\frac{||A||_1}{N^2}$. For Eq.6, the discriminator $G$ only needs to be able to determine whether the patches are homologous. That is, the affinity of different source patches is suppressed, and the affinity of homologous patches is improved.

\begin{figure}[ht]
	\centering
	\includegraphics[width=\linewidth,scale=1.00]{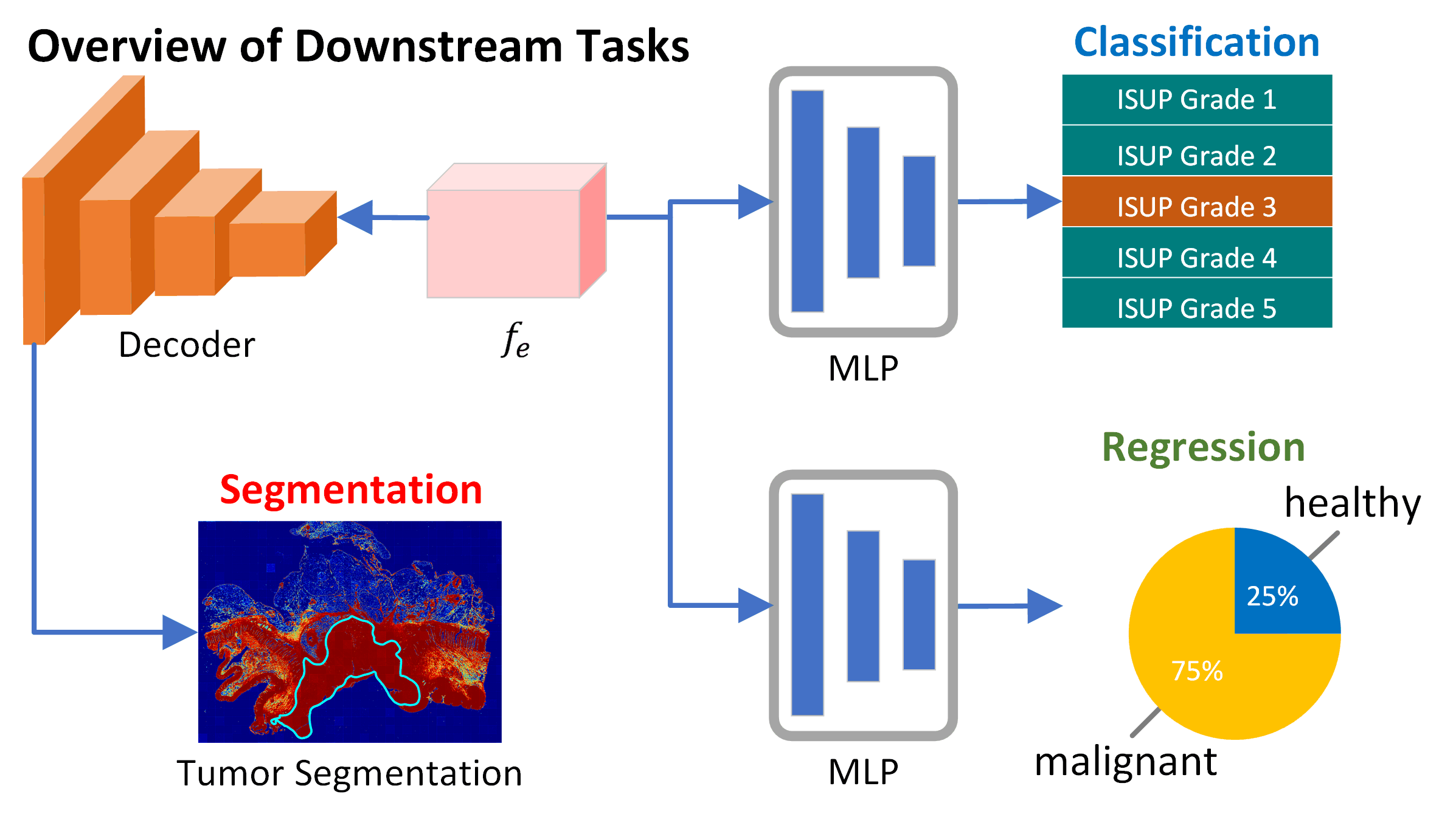}
	\caption{Overview of downstream tasks of pathological image analysis.}
	\label{Fig.4}
\end{figure}

\subsection{Downstream Task Migration}
\label{Chapter 3.3}
The backbone after the above self-supervised training has a good representation extraction ability. Then, after the downstream tasks, the learned parameters can be migrated and fine-tuned, just like the general supervised learning process. Therefore, the ability of self-supervised learning is mainly reflected by the performance of downstream tasks. As shown in step 2 of Fig.1, we improve the general self-supervised migration process by introducing residual learning and attention mechanisms. The parameters of the generic feature extractor $E_{e_g}$ after self-supervised learning are fixed, the gradient is turned off, and the features extracted by it are regarded as generic features. The parallel special feature extractor $E_{e_s}$ obtains the special feature $f_{e_s}$ by residual learning on the basis of the generic feature $f_{e_g}$. Specifically, it can be expressed as the following process:

\begin{equation}
	f_{e_g}^l = E_{e_g}^l(f_{e_g}^{l-1}),
\end{equation}

\begin{equation}
	f_{e_s}^l = E_{e_s}^l(f_{e_s}^{l-1}) + f_{e_g}^l,
\end{equation}

Where, $f_{e_g}^l$ is the general feature of the upper layer feature $f_{e_g}^{l-1}$ through the $l$ layer of $E_{e_g}^l$. $f_{e_g}^l$ is the special feature the upper layer feature $f_{e_s}^{l-1}$ through the $l$ layer of $E_{e_s}^l$ and add the residual item $f_{e_g}^l$. $f_{e_g}^l$ of the same layer can be regarded as the feature modification of $f_{e_s}^l$, that is, the special feature is the result of residual learning on the available feature. For the final general features $f_{e_g}$ and special features $f_{e_s}$, we combined them in the form of attentional mechanisms get $f_{e}$:
\begin{equation}
	f_{e} = f_{e_g} * f_{e_s} + \sigma(f_{e_s}),
\end{equation}

Next, $f_{e}$ can be easily embedded into the pathological image analysis downstream task introduced in related work Chapter 2. As shown in Fig.4, we briefly summarize this paper's three forms of downstream tasks experiment. In the classification task, $f_{e}$ inputs MLP to obtain each category's confidence, and Softmax gets the classification result. In Fig.4, we illustrate the classification of ISUP Grade in breast cancer. Similarly, in the regression task, $f_{e}$ also inputs MLP and finally inputs the nonlinear layer to output logistic regression results. In Fig.4, we illustrate the regression of the proportion of malignant and healthy cells. In the segmentation task, after $f_{e}$ is input to the decoder, it is recovered and decoded layer by layer through deconvolution to generate the segmentation mask. In Fig.4, we exemplify tumor segmentation.

\section{Experiment}

\subsection{Datasets Preparation}
Our experiments are mainly divided into two parts. First, we compared SASSL with other staining-normalized SOTA models. Secondly, we compared SASSL with other self-supervised methods for different pathological image downstream analysis tasks to thoroughly verify the adaptive ability and transfer robustness of SASSL to the staining changes of pathological slides. In the first part, We use a selection of histopathological datasets with 200 WSIs from 4 different tumors, 1) colon adenocarcinoma from TCGA-COAD(\cite{tcgacoda}), 2) rectum adenocarcinoma from TCGA-READ(\cite{tcgacoda}), 3) stomach adenocarcinoma from TCGA-STAD(\cite{tcgacoda}) and 4) breast cancer metastases from Camelyon16(\cite{GeertPeter2018}), with a balanced distribution of 50 slides each. This part of WSIs were split into non-overlapped 224 × 224 patches at the magnitude of 20× to retain the high resolution. In the Second part, We evaluate the performance of our method on four pathological image analysis datasets: PANDA(\cite{Bjartell2005}) data sets, BreastPathQ(\cite{MohammadPeikari2017})data sets,CAMELYON16(\cite{GeertPeter2018})data sets. Among the PANDA data sets is the classification data set, the BreastPathQ data set is the regression data set, and the CAMELYON16 data sets are segmentation data sets. This part of WSIs were split into non-overlapped 512 × 512 patches at the magnitude of 20× to retain the high resolution. All the datasets are publicly available online.

\textit{TCGA-COAD}(\cite{tcgacoda})\footnote{\href{https://wiki.cancerimagingarchive.net/pages/viewpage.action?pageId=16712033}{TCGA-COAD}}. The Cancer Genome Atlas Colon Adenocarcinoma (TCGA-COAD) data collection is part of a larger effort to build a research community focused on connecting cancer phenotypes to genotypes by providing clinical images matched to subjects from The Cancer Genome Atlas (TCGA).The TCGA-COAD Data set consists of 983 WSIs of digitized H{\&}E-stained. We divided 70 \% of the data into training sets and 30\% into test sets. We split each WSI cut into patches at 20x magnification with a size of $224\times224$ with a balanced distribution of 50 slides each. we split into 49150 patches, the overall label for subordinate patches obtained by each WSIs.

\textit{TCGA-READ}(\cite{tcgacoda})\footnote{\href{https://wiki.cancerimagingarchive.net/pages/viewpage.action?pageId=16711737}{TCGA-READ}}. The Cancer Genome Atlas Rectum Adenocarcinoma (TCGA-READ) data collection is part of a larger effort to build a research community focused on connecting cancer phenotypes to genotypes by providing clinical images matched to subjects from The Cancer Genome Atlas (TCGA). The TCGA-COAD Data set consists of 364 WSIs of digitized H{\&}E-stained. We divided 70 \% of the data into training sets and 30\% into test sets. We split each WSI cut into patches at 20x magnification with a size of $224\times224$ with a balanced distribution of 50 slides each. we split into 18200 patches, the overall label for subordinate patches obtained by each WSIs.

\textit{TCGA-STAD}(\cite{tcgacoda})\footnote{\href{https://wiki.cancerimagingarchive.net/pages/viewpage.action?pageId=19039400}{TCGA-STAD}}.The Cancer Genome Atlas Stomach Adenocarcinoma (TCGA-STAD) data collection is part of a larger effort to build a research community focused on connecting cancer phenotypes to genotypes by providing clinical images matched to subjects from The Cancer Genome Atlas (TCGA). The TCGA-COAD Data set consists of 758 WSIs of digitized H{\&}E-stained. We divided 70 \% of the data into training sets and 30\% into test sets. We split each WSI cut into patches at 20x magnification with a size of $224\times224$ with a balanced distribution of 50 slides each. we split into 37900 patches, the overall label for subordinate patches obtained by each WSIs.

\textit{PANDA}(\cite{Bjartell2005})\footnote{\href{https://www.kaggle.com/c/prostate-cancer-grade-assessment}{PANDA}}. Prostate cancer (PCa) is the second most common cancer in men in the world. Pathologists score the diagnosis based on the Gleason scoring system. In the Gleason scoring system, the rating is converted to an ISUP rating of 1-5. the ISUP grading plays a key role in determining how patients should be treated.The PANDA Data set consists of around 10,616 WSIs of digitized H{\&}E-stained from two centers. There were 5,456 from Karolinska and 5,160 from radboud. We divided 80 \% of the data into training sets and 20\% into test sets. We split each WSI cut into patches at 20x magnification with a size of $512\times512$. we split into 406,894 patches, the overall label for subordinate patches obtained by each WSIs is the ISUP.

\textit{BreastPathQ}(\cite{MohammadPeikari2017})\footnote{\href{https://breastpathq.grand-challenge.org/}{BreastPathQ}}. Breast cancer is the phenomenon of runaway proliferation of mammary epithelial cells under the action of various carcinogenic factors. Often referred to as the "pink killer," breast cancer is the number one malignancy in women,The assessment of cell structure is an important part of tumor load assessment.The number of cells in the tumor bed is defined as the percentage area of the entire tumor bed composed of tumor cells (invasive or in situ).In the BreastPathQ data set, 94 WSIs (69 training and 25 testing) that had been stained with H{\&}E-stained were included, and 3700 ROI patches were extracted from the above WSI and marked with tumor cell fractions (0-100\%) in these areas. The training set contains ROI patches 2579, the test set contains ROI patches 1121, and each patch is $512\times512$ in size. 

\textit{CAMELYON16}(\cite{GeertPeter2018})\footnote{\href{https://camelyon16.grand-challenge.org/}{CAMELYON16}}. Automatic detection of lymph node metastasis has great potential to help pathologists and reduce their workload. The purpose of CAMELYON16 data set is to realize the automatic detection of cancer metastasis in lymph node images. CAMELYON16 include 400 hematoxylin-eosin (H\&E) stained whole-slide images (WSIs) of lymph node sections. Among the 400 WSI sections, 270 WSIs (110 contain cancerous) were training sets and the remaining 130 WSIs (49 contain cancerous) were test sets. Even in cancer WSIs, normal tissues still account for a larger proportion. If too many normal WSIs are introduced for training, the model will not be able to learn the characteristics of cancer tissues well. Therefore, we only used WSIs containing cancer tissues for training and testing. For the first part of the experiment, We split each cancerous WSI cut into patches at 20x magnification with a size of $224\times224$. we split into 13500 patches in the train set, On the test set, we split into 6500 patches.For the second part of the experiment, we split each cancerous WSI cut into patches at 20x magnification with a size of $512\times512$. we split into 7853 patches in the train set, On the test set, we use the complete WSIs of the test set to evaluate the segmentation accuracy of the model.

In self-supervised training, We randomly selected 1000 WSIs from three data sets and filtered the background patches through OTSU. For each WSIs, we sampled 100 foreground patches in a balanced manner to ensure data balance between different stains. In the downstream task, the data set is segmented and trained according to the above description to verify the algorithm's effectiveness. Therefore, we get a self-supervised training set containing $10^5$ patches and use this training set for the self-supervised training. It is worth noting that in the self-monitoring training process, when the batch size is n, we balanced the number of patches belonging to the same slice, so as to avoid the situation that there are no homologous patches in the training process, that is, the relationship matrix is Fig3a.

\begin{table*}[ht]
	\label{Table1}
	\caption{\newline Comparison of the Sate-Of-The-Art of stain normalization benchmark in the four data test set. We use $Acc$(Accuracy) as a metric to evaluate classification performances with SASSL and SOTA methods. The last column is the average performance of the different models across the four data sets.}
	\centering
	\begin{tabular}{c|c|c|c|c|c}
		\hline
		Method & $Camelyon16$ & $TCGA-STAD$ & $TCGA-COAD$ & $TCGA-READ$ & $Aggregrated$\\
		\hline
		\cite{Reinhard} & 0.893 & 0.901  & 0.903  & 0.883 & 0.895 \\
		\cite{Macenko} & 0.867 & 0.893 & 0.912 & 0.905 & 0.894 \\
		\cite{Khan} & 0.912 & 0.902 & 0.893 & 0.911 & 0.904 \\
		\cite{Vahadane} & 0.908 & 0.914 & 0.908 & 0.906 & 0.909 \\
		\cite{Staingan} & 0.918 & 0.921 & 0.922 & 0.912 & 0.918 \\
		\cite{Harshal} & 0.913 & 0.904 & 0.912 & 0.907 & 0.909 \\ 
		\cite{Saad} [11] & 0.878 & 0.902 & 0.894 & 0.889 & 0.891 \\ 
		\cite{KeJing} & 0.923 & 0.927 & 0.921 & 0.925 & 0.924 \\
		\hline
		EfficientNet-$B_0$ & 0.881 & 0.893 & 0.886 & 0.886 & 0.886 \\
		EfficientNet-$B_0$(SASSL) & 0.909 & 0.914 & 0.917 & 0.919 & 0.914 \\ 
		\hline
		EfficientNet-$B_1$ & 0.911 & 0.923 & 0.915 & 0.917 & 0.916 \\
		EfficientNet-$B_1$(SASSL) & 0.942 & 0.952 & 0.953 & 0.957 & 0.951 \\ 
		\hline
		ResNet34 & 0.876 & 0.881 & 0.884 & 0.882 & 0.881 \\
		ResNet34(SASSL) & 0.908 & 0.921 & 0.925 & 0.917 & 0.918 \\ 
		\hline
		ResNet50 & 0.864 & 0.886 & 0.874 & 0.879 & 0.876 \\
		ResNet50(SASSL) & 0.926 & 0.944 & 0.935 & 0.932 & 0.934 \\ 
		\hline
		RepVGG16 & 0.894 & 0.903 & 0.902 & 0.908 & 0.901 \\
		RepVGG16(SASSL) & 0.927 & 0.925 & 0.928 & 0.925 & 0.923 \\ 
		\hline
		RepVGG19 & 0.917 & 0.909 & 0.909 & 0.911 & 0.911 \\
		RepVGG19(SASSL) & 0.926 & 0.931 & 0.928 & 0.931 & 0.929 \\ 
		\hline
		VIT & 0.852 & 0.854 & 0.88 & 0.855 & 0.858 \\
		VIT(SASSL) & 0.887 & 0.89 & 0.881 & 0.875 & 0.883 \\ 
		\hline
		Swin & 0.855 & 0.868 & 0.867 & 0.867 & 0.864 \\
		Swin(SASSL) & 0.893 & 0.905 & 0.893 & 0.887 & 0.894 \\ 
		\hline
		NASNet & 0.864 & 0.873 & 0.868 & 0.871 & 0.869 \\
		NASNet(SASSL) & 0.897 & 0.901 & 0.899 & 0.895 & 0.898 \\ 
		\hline
		ProxylessNAS & 0.886 & 0.897 & 0.897 & 0.897 & 0.894 \\
		ProxylessNAS(SASSL) & 0.904 & 0.922 & 0.912 & 0.923 & 0.915 \\ 
		\hline
	\end{tabular} 
\end{table*}

\begin{table*}[ht]
	\caption{\newline Comparison of the structural ablation of different self-supervised frameworks base on ResNet50 in the PANDA data set. In PANDA Dataset testing set got the following average scores. $QWK$ means Quadratic weighted kappa. $Acc$ means Accuracy score. $F_1^0-F_1^5$ represents the scores of different ISUP categories, and $F_1^{micro}$ means the micro F1 score of all categories.}
	\label{Table2}
	\centering
	\begin{tabular}{c|c|c|c|c|c|c|c|c|c|c|c}
		\toprule[1pt]
		Method & $QWK_{all}$ & $QWK_{k}$ & $QWK_{r}$ & $Acc$ & $F_1^0$ & $F_1^1$ & $F_1^2$ & $F_1^3$ & $F_1^4$ & $F_1^5$ & $F_1^{micro}$ \\
		\toprule[1pt]
		BYOL & 0.878 & 0.846 & 0.865 & 0.598 & 0.810 & 0.643 & 0.444 & 0.387 & 0.468 & 0.542 & 0.549\\
		BYOL(SASSL, w/o RL) & 0.898 & 0.866 & 0.891 & \textbf{0.62} & 0.799 & 0.650 & \textbf{0.484} & \textbf{0.451} & 0.527 & 0.6 & 0.585\\ 
		BYOL(SASSL, w/ RL) & \textbf{0.907} & \textbf{0.887} & \textbf{0.894} & 0.453 & \textbf{0.860} & \textbf{0.668} & 0.467 & 0.444 & \textbf{0.538} & \textbf{0.640} & \textbf{0.603} \\ 
		\hline
		SimSiam & 0.872 & 0.856 & 0.856 & 0.578 & 0.753 & 0.622 & 0.454 & 0.407 & 0.498 & 0.515 & 0.541\\ 
		SimSiam(SASSL, w/o RL) & 0.903 & 0.891 & 0.894 & 0.641 & 0.837 & \textbf{0.686} & \textbf{0.459} & \textbf{0.441} & 0.532 & 0.593 & 0.592\\ 
		SimSiam(SASSL, w/ RL) & \textbf{0.910} & \textbf{0.903} & \textbf{0.896} & \textbf{0.645} & \textbf{0.842} & 0.682 & 0.438 & 0.407 & \textbf{0.543} & \textbf{0.696} & \textbf{0.601}\\ 
		\hline
		SimCLR & 0.862 & 0.845 & 0.837 & 0.566 & 0.748 & 0.607 & 0.456 & 0.416 & 0.425 & 0.493 & 0.524\\ 
		SimCLR(SASSL, w/o RL) & 0.877 & 0.860 & \textbf{0.858} & 0.608 & \textbf{0.829} & 0.647 & 0.443 & 0.425 & 0.441 & 0.553 & 0.556\\ 
		SimCLR(SASSL, w/RL) & \textbf{0.890} & \textbf{0.873} & 0.708 & \textbf{0.639} & 0.808 & \textbf{0.709} & \textbf{0.480} & \textbf{0.474} & \textbf{0.487} & \textbf{0.573} & \textbf{0.588}\\ 
		\hline
		MoCo & 0.865 & 0.844 & 0.839 & 0.591 & 0.782 & 0.634 & 0.478 & 0.442 & 0.432 & 0.5 & 0.545\\ 
		MoCo(SASSL, w/o RL) & 0.883 & 0.865 & 0.865 & 0.610 & 0.803 & 0.656 & 0.494 & \textbf{0.459} & 0.449 & 0.525 & 0.564\\  
		MoCo(SASSL, w/ RL) & \textbf{0.893} & \textbf{0.875} & \textbf{0.875} & \textbf{0.624} & \textbf{0.806} & \textbf{0.691} & \textbf{0.498} & 0.439 & \textbf{0.457} & \textbf{0.589} & \textbf{0.580}\\ 
		\hline
	\end{tabular} 
\end{table*}

\begin{table*}[ht]
	\qquad
	\begin{minipage}[t]{0.45\textwidth}
		\centering
		\caption{\newline Comparison of the structural ablation of different self-supervised frameworks base on ResNet50 in the BreastPathQ test data set. $MAE$ means Mean Absolute error. $MSE$ means Mean Square Error, and $R^2$ means the R-Square.}
		\begin{tabular}{c|c|c|c}
			\hline
			Method & $MAE$ & $MSE$ & $R^2$ \\
			\hline
			BYOL & 0.198 & 0.060 & 0.425 \\
			BYOL(SASSL, w/o RL) & 0.190 & 0.058 & 0.441 \\ 
			BYOL(SASSL, w/ RL) & \textbf{0.186} & \textbf{0.055} & \textbf{0.474} \\ 
			\hline
			SimSiam & 0.187 & 0.056 & 0.468 \\
			SimSiam(SASSL, w/o RL) & 0.181 & 0.052 & 0.504 \\ 
			SimSiam(SASSL, w/ RL) & \textbf{0.179} & \textbf{0.051} & \textbf{0.512} \\ 
			\hline
			SimCLR & 0.195 & 0.060 & 0.431 \\
			SimCLR(SASSL, w/o RL) & 0.187 & 0.056 & 0.468 \\ 
			SimCLR(SASSL, w/ RL) & \textbf{0.180} & \textbf{0.051} & \textbf{0.508} \\
			\hline
			MoCo & 0.191 & 0.058 & 0.445 \\
			MoCo(SASSL, w/o RL) & 0.188 & \textbf{0.056} & \textbf{0.461} \\ 
			MoCo(SASSL, w/ RL) & \textbf{0.186} & 0.059 & 0.451 \\
			\hline
		\end{tabular}
		\label{Table3}
	\end{minipage}
	\qquad
	\begin{minipage}[t]{0.45\textwidth}
		\centering
		\caption{\newline Comparison of the structural ablation of different self-supervised frameworks base on ResNet50 in the CAMELYON16 test data set.$PA$ means Pixel Accuracy, $Dice$ means Dice coefficient and $MIoU$ means mean Intersection over Union}
		\begin{tabular}{c|c|c|c}
			\toprule[1pt]
			Method & $PA$ & $Dice$ & $MIoU$ \\
			\toprule[1pt]
			BYOL & 0.842 & 0.816 & 0.724  \\
			BYOL(SASSL, w/o RL) & 0.874 & 0.823 & 0.734 \\ 
			BYOL(SASSL, w/ RL) & \textbf{0.893} & \textbf{0.835} & \textbf{0.758} \\ 
			\hline
			SimSiam & 0.905 & 0.840 & 0.766 \\
			SimSiam(SASSL, w/o RL) & 0.912 & 0.872 & 0.774 \\ 
			SimSiam(SASSL, w/ RL) & \textbf{0.929} & \textbf{0.906} & \textbf{0.839} \\
			\hline
			SimCLR & 0.886 & 0.842 & 0.728 \\
			SimCLR(SASSL, w/o RL) & 0.903 & 0.891 & \textbf{0.805} \\ 
			SimCLR(SASSL, w/ RL) & \textbf{0.912} & \textbf{0.902} & 0.782 \\
			\hline
			MoCo & 0.860 & 0.909 & 0.843 \\
			MoCo(SASSL, w/o RL) & 0.886 & 0.927 & \textbf{0.826} \\ 
			MoCo(SASSL, w/ RL) & \textbf{0.894} & \textbf{0.936} & 0.823 \\
			\hline
		\end{tabular}
		\label{Table4}
	\end{minipage}
	
\end{table*}

\begin{figure*}[ht]
	\centering
	\includegraphics[width=0.8\linewidth,scale=1.00]{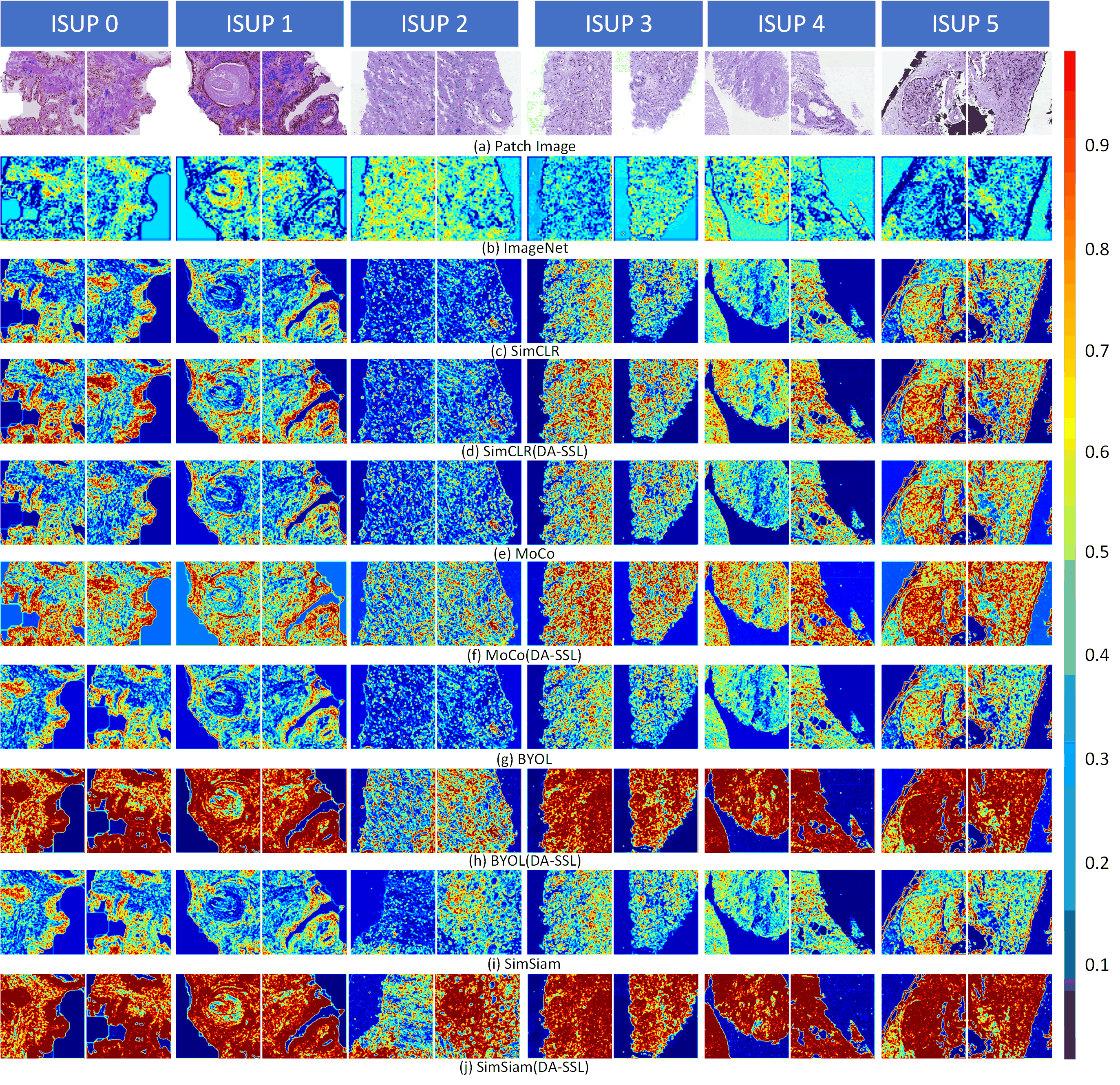}
	\caption{Panda DataSet HeatMap with Differnent Self-Supervisor.}
	\label{Fig.6}
\end{figure*}

\begin{table*}[ht]
	\label{Table5}
	\caption{\newline Comparison of the structural ablation of different classifier baseline in the PANDA data set. In PANDA Dataset Testing set got the following average scores. $QWK$ means Quadratic weighted kappa. $Acc$ means Accuracy score. $F_1^0-F_1^5$ represents the scores of different ISUP categories, and $F_1^{micro}$ means the micro F1 score of all categories.}
	\centering
	\begin{tabular}{c|c|c|c|c|c|c|c|c|c|c|c}
		\hline
		Method & $QWK_{all}$ & $QWK_{k}$ & $QWK_{r}$ & $Acc$ & $F_1^0$ & $F_1^1$ & $F_1^2$ & $F_1^3$ & $F_1^4$ & $F_1^5$ & $F_1^{micro}$ \\
		\hline
		EfficientNet-$B_0$ & 0.868 & 0.832 & 0.854 & 0.603 & 0.799 & 0.677 & 0.411 & 0.412 & 0.419 & 0.522 & 0.540\\
		EfficientNet-$B_0$(SASSL) & 0.898 & 0.866 & 0.891 & 0.62 & 0.799 & 0.650 & 0.484 & 0.451 & 0.527 & 0.6 & 0.585\\ 
		\hline
		EfficientNet-$B_1$ & 0.905 & 0.887 & 0.883 & 0.618 & 0.839 & 0.705 & 0.472 & 0.457 & 0.502 & 0.608 & 0.597\\ 
		EfficientNet-$B_1$(SASSL) & \textbf{0.935} & \textbf{0.917} & \textbf{0.923} & \textbf{0.701} & \textbf{0.885} & \textbf{0.761} & \textbf{0.526} & \textbf{0.505} & \textbf{0.534} & \textbf{0.681} & \textbf{0.649} \\ 
		\hline
		ResNet34 & 0.832 & 0.795 & 0.801 & 0.496 & 0.640 & 0.543 & 0.378 & 0.340 & 0.330 & 0.594 & 0.471\\ 
		ResNet34(SASSL) & 0.871 & 0.846 & 0.850 & 0.600 & 0.772 & 0.66 & 0.457 & 0.407 & 0.440 & 0.568 & 0.550\\ 
		\hline
		ResNet50 & 0.872 & 0.856 & 0.856 & 0.578 & 0.753 & 0.622 & 0.454 & 0.407 & 0.498 & 0.515 & 0.541\\ 
		ResNet50(SASSL) & 0.910 & 0.903 & 0.896 & 0.645 & 0.842 & 0.682 & 0.438 & 0.407 & 0.543 & 0.696 & 0.601\\ 
		\hline
		RepVGG16 & 0.743 & 0.708 & 0.676 & 0.426 & 0.447 & 0.485 & 0.375 & 0.373 & 0.430 & 0.330 & 0.407\\ 
		RepVGG16(SASSL) & 0.756 & 0.719 & 0.735 & 0.578 & 0.753 & 0.622 & 0.454 & 0.407 & 0.498 & 0.515 & 0.541\\ 
		\hline
		RepVGG19 & 0.773 & 0.702 & 0.753 & 0.432 & 0.374 & 0.551 & 0.357 & 0.352 & 0.402 & 0.342 & 0.396\\ 
		RepVGG19(SASSL) & 0.768 & 0.732 & 0.700 & 0.460 & 0.582 & 0.522 & 0.381 & 0.360 & 0.350 & 0.380 & 0.429\\ 
		\toprule[1pt]
		VIT & 0.755 & 0.735 & 0.671 & 0.477 & 0.394 & 0.599 & 0.451 & 0.517 & 0.277 & 0.204 & 0.418\\ 
		VIT(SASSL) & 0.799 & 0.709 & 0.715 & 0.475 & 0.414 & 0.568 & 0.449 & 0.493 & 0.453 & 0.209 & 0.431\\ 
		\hline
		Swin & 0.793 & 0.772 & 0.743 & 0.573 & 0.465 & 0.709 & \textbf{0.529} & \textbf{0.602} & 0.476 & 0.537 & 0.525\\ 
		Swin(SASSL) & \textbf{0.833} & \textbf{0.806} & \textbf{0.779} & \textbf{0.607} & \textbf{0.682} & \textbf{0.716} & 0.493 & 0.586 & \textbf{0.506} & \textbf{0.646} & \textbf{0.555}\\ 
		\toprule[1pt]
		NASNet & 0.830 & 0.768 & 0.825 & 0.554 & 0.700 & 0.586 & 0.462 & 0.398 & 0.454 & 0.561 & 0.527\\ 
		NASNet(SASSL) & 0.861 & 0.813 & 0.851 & 0.527 & 0.768 & 0.508 & 0.384 & 0.308 & 0.387 & 0.578 & 0.489\\ 
		\hline
		ProxylessNAS & 0.855 & 0.815 & 0.844 & \textbf{0.593} & 0.710 & \textbf{0.650} & \textbf{0.499} & \textbf{0.436} & \textbf{0.485} & 0.584 & \textbf{0.561}\\ 
		ProxylessNAS(SASSL) & \textbf{0.882} & \textbf{0.853} & \textbf{0.869} & 0.578 & \textbf{0.756} & 0.599 & 0.458 & 0.410 & 0.473 & \textbf{0.606} & 0.550\\ 
		\hline
	\end{tabular} 
\end{table*}

\subsection{Evaluation Metric}

\begin{table}[ht]
	\centering
	\caption{\newline Comparison of the structural ablation of different different regressor baseline in the BreastPathQ data set.}
	\begin{tabular}{c|c|c|c}
		\toprule[1pt]
		Method & $MAE$ & $MSE$ & $R^2$ \\
		\toprule[1pt]
		EfficientNet-$B_0$ & 0.198 & 0.060 & 0.425 \\
		EfficientNet-$B_0$(SASSL) & 0.190 & 0.058 & 0.441 \\ 
		\hline
		EfficientNet-$B_1$ & 0.187 & 0.056 & 0.468 \\
		EfficientNet-$B_1$(SASSL) & 0.181 & \textbf{0.052} & 0.504 \\ 
		\hline
		ResNet34 & 0.191 & 0.058 & 0.448 \\
		ResNet34(SASSL) & 0.183 & 0.053 & 0.491 \\ 
		\hline
		ResNet50 & 0.187 & 0.056 & 0.468 \\
		ResNet50(SASSL) & \textbf{0.179} & 0.051 & \textbf{0.512} \\ 
		\hline
		RepVGG16 & 0.225 & 0.077 & 0.425 \\
		RepVGG16(SASSL) & 0.204 & 0.066 & 0.371 \\ 
		\hline
		RepVGG19 & 0.205 & 0.067 & 0.363 \\
		RepVGG19(SASSL) & 0.187 & 0.058 & 0.441 \\ 
		\toprule[1pt]
		VIT & 0.199 & 0.063 & 0.399 \\
		VIT(SASSL) & \textbf{0.181} & \textbf{0.052} & \textbf{0.504} \\ 
		\hline
		Swin & 0.235 & 0.083 & 0.212 \\
		Swin(SASSL) & 0.209 & 0.067 & 0.357 \\ 
		\toprule[1pt]
		NASNet & 0.195 & 0.060 & 0.431 \\
		NASNet(SASSL) & \textbf{0.187} & \textbf{0.056} & \textbf{0.468} \\ 
		\hline
		ProxylessNAS & 0.191 & 0.058 & 0.445 \\
		ProxylessNAS(SASSL) & 0.188 & \textbf{0.056} & 0.461 \\ 
		\toprule[1pt]
	\end{tabular}
	\label{Table6}
\end{table}

In the first part, We use $Acc$(Accuracy) as a metric to evaluate classification performances with SASSL and SOTA methods. In the second part, We use $WQK$(Weighted Quadratic Kappa), $F_1$($F_1$ Score), and $Acc$(Accuracy) as a metric to evaluate classification performances. Weighted Quadratic Kappa measures the agreement between two outcomes. This metric typically varies from 0 (random agreement) to 1 (complete agreement). If there is less agreement than expected by chance, the metric may go below 0. The quadratic weighted kappa is calculated as follows. First, an $N \times N$ histogram matrix $O$ is constructed, such that $O_{i j}$ corresponds to the number of categories $i$ (actual) that received a predicted value value $j$. An $N$-by-$N$ matrix of weight $w$ is calculated based on the difference between actual and predicted values: $w_{ij}=\frac{(i-j)^2}{(N-1)^2}$. An $N$-by-$N$  histogram matrix of expected outcomes $E$ is calculated, assuming no correlation exists between values. This is calculated as the outer product between the actual histogram vector of outcomes and the predicted histogram vector, normalized such that $E$ and $O$ have the same sum. From these three matrices, the quadratic weighted kappa is calculated as:
\begin{equation}
	kappa = 1 - \frac{\sum_{ij}w_{ij}O_{ij}}{\sum_{ij}w_{ij}E_{ij}},
\end{equation}
And then also, $F_1$($F_1$ Score) and $Acc$(Accuracy) was introduced as the evaluation criterion used to evaluate the classification result. We use $MSE$, $MAE$, and $R^2$ to verify regression performance. Finally, $PA$(Pixel Accuracy), $Dice$ coefficient, and $MIoU$(mean Intersection over Union) were used to evaluate the segmentation performance. 

\section{Results And Discussion}

\subsection{Comparison to state-of-the-art stain normalization frameworks}
In the Frist part of experiment, We evaluated the proposed method with a couple of Sate-Of-The-Art(SOTA) stain normalization approaches proposed by \cite{Reinhard}, \cite{Macenko}, \cite{Khan}, \cite{Vahadane}, \cite{Staingan}, \cite{Harshal}, and \cite{Saad}. The demonstrated results are performed on test sets. We use the EfficientNet(\cite{TanLe2019}), ResNet(\cite{HeZhang2015}), RepVGG(\cite{DingZhang2021}), VIT(\cite{DosovitskiyBeyer2021}), Swin(\cite{LiuLin2021}), NASNet(\cite{ZophVasudevan2018}), and ProxylessNAS(\cite{CaiZhu2019}) models as the benchmark backbone network, where EfficientNet(\cite{TanLe2019}), ResNet(\cite{HeZhang2015}), and RepVGG(\cite{DingZhang2021}) are currently representative work based on the results of convolutional neural networks. VIT(\cite{DosovitskiyBeyer2021}) and Swin(\cite{LiuLin2021}) are currently novel transformer structures, and NASNet(\cite{ZophVasudevan2018}) and ProxylessNAS(\cite{CaiZhu2019}) are based on the network structure search framework. These three models represent the current backbone network design benchmarks, and we hope to compare the performance changes after adding SASSL to these three models to verify their effectiveness of SASSL. The experimental results show in Table 1. It can be seen from the experimental results that when the model after adding SASSL, the classification accuracy has been steadily improved, and the improvement range is between 1\% and 6\%. The results of each model on the four datasets have significantly improved, making the model perform better than the results without SASSL on average. Compared with the general staining-standardized SOTA method, the improvement of SASSL is more significant. As shown in Table 1, SASSL outperforms the best SOTA stain normalization method \cite{KeJing} by 2.7\% on EfficientNet-B1. The results of SASSL are simpler and clearer and can be more widely used in pathological image analysis tasks. The reason is because SASSL is not limited to the general idea of forced alignment based on dye migration from the perspective of dye adaptation. Instead, it tries to align the semantic features of pathological images in a self-supervised form to improve the model's ability to express the key features of pathological images while achieving self-adaptive staining. SASSL also shows better staining adaptability and feature extraction robustness than other staining normalization methods.

\subsection{Performance of structural ablation on different self-supervised frameworks}

Next, We verified the effects of our proposed SASSL method on different self-supervised frameworks. We selected SimCLR(\cite{ChenKornblith2020v1}), MoCo(\cite{ChenFan2020}), BYOL(\cite{GrillStrub2020}), and SimSiam(\cite{ChenHe2021}) as benchmarks. Among them, SimCLR(\cite{ChenKornblith2020v1}) and MoCov2(\cite{ChenFan2020})(Fig.3a) belong to the self-supervised paradigm based on positive and negative samples, while BYOL(\cite{GrillStrub2020}) and SimSiam(\cite{ChenHe2021})(Fig.3b) belong to the self-supervised paradigm based on positive samples.We conducted comparative experiments on three data sets. The distribution of these three data sets belongs to different types of downstream tasks. The results of the PANDA data set are shown in Table 2, the results of the BreastPathQ data set are shown in Table 3, and the results of the CAMEONLY16 data set are shown in Table 4. In Table 1, we used ResNet50 as the backbone to perform the PANDA testing dataset score. In the method column, we compared the changes in the four self-supervised frameworks after the introduction of SASSL. $QWK$ means Quadratic weighted kappa. $ACC$ means Accuracy scores. $F_1^0-F_1^5$ represents the scores of different ISUP categories, and $F_1^{micro}$ means the micro $F_1$ score of all categories. It can be seen from the experimental results that SASSL can effectively improve downstream classification performance regardless of whether residual learning is performed. It can be seen from the QWK score transformation that the SASSL method not only improves the overall QWK score but also improves the classification accuracy of data from different central sources. This further shows that SASSL can significantly improve the model's resistance to staining differences. For the four different self-supervised frameworks, QWK is improved by 1\%-2\% after using the SASSL method, and the overall improvement is between 2\%-4\% after the introduction of residual training. It is worth noting that in addition to robust improvement of samples from different data sources, SASSL can also generally improve the F1 value of each category.

In order to further analyze the improvement of the model feature learning ability by SASSL, we used CAM(\cite{ZhouKhosla2016}) to visualize the calorific value maps of different self-supervised frameworks after using SASSL for comparison, As shown in Fig.5. In most current methods, the parameter initialization of the model is to use ImageNet's pre-trained model weights. This can also bring a certain degree of improvement in the general scene data set. However, due to the significant difference between the characteristics of the medical image and the scene image, the advantage of this method in many medical image analysis works is not apparent. It can be seen in (Fig 5a) that the heating value effect obtained by initializing the weights of the ImageNet pre-training model is insignificant. The model pays more attention to the superficial features of pathological images, and there is no distinction between areas with apparent differences. Although the common self-supervised framework (Fig 5c,e,g, i) can effectively distinguish the critical areas in the pathological image, it has a higher calorific value corresponding to the tissue area. However, some areas with significant differences in dyeing still cannot produce higher heating values. After using SASSL  (Fig 5d,f,h,j), common self-monitoring frameworks have a higher calorific response to key organizational areas. Among them, the improvement of the self-supervision method based on positive samples is the most obvious. It can be seen that the model has a high response to the tissue area and can effectively distinguish the unorganized area. After using SASSL for self-supervised training, downstream models can quickly locate key tissue regions and distinguish the difference between lesions and normal tissues.

In addition to classification tasks, SASSL is also effective in regression and segmentation tasks, as shown in Table 2. After adding SASSL for different self-supervised frameworks, the R2 of downstream models generally rises by 2\%-4\%. After the introduction of residual learning, it generally rises by 1\%-7\%, proving that SASSL is effective for downstream models. In the segmentation task, the result of CAMELYON16 is shown in Table 3. For different self-supervised frameworks, after adding SASSL, the R2 of downstream models generally rises by 1\%-3\%, and after the introduction of residual learning, it generally rises by 1\%-6\%. It can be seen from the results of use in Table 1-3 that SASSL is effective in different self-supervised methods. Furthermore, suitable for many types of tasks, SASSL can produce stable improvements.

\subsection{Performance of structural ablation on different backbone}

In previous experiments, we have demonstrated the effectiveness of SASSL on different self-supervised frameworks. In order to further prove the generalization of SASSL, we selected corresponding benchmark models for different task types, and we introduced SASSL for these benchmark models to analyze the changes in model performance. In order to make the experimental results more intuitive and avoid redundancy, the previous experimental results are integrated. Furthermore, the previous experimental results proved that the introduction of residual learning is also stable for the improvement of the model, so we selected the most stable performance of SimSiam as the self-supervised framework benchmark and performed residual learning. Under this benchmark, a unified comparison experiment was carried out. We all used SimSiam for self-supervised learning for the original SOTA model, which was compared with the addition of SASSL and residual learning.

\textit{Classification tasks}.We use the EfficientNet(\cite{TanLe2019}), ResNet(\cite{HeZhang2015}), RepVGG(\cite{DingZhang2021}), VIT(\cite{DosovitskiyBeyer2021}), Swin(\cite{LiuLin2021}), NASNet(\cite{ZophVasudevan2018}), and ProxylessNAS(\cite{CaiZhu2019}) models as the classification benchmark. The experimental results are shown in Table 5. From the results, it can be seen that SASSL has significantly improved models based on the Transformer structure, such as VIT and Swin, and QWK has increased by 4.4\% and 4\%, respectively. For other types of models, the QWK of the model increased by 1\%-4\% after joining SASSL. Among them, the best performance is EfficientNet-B1, QWK reaches 0.935, which is an increase of 6\% compared to the simple application of SimSiam. Other SOTA models have significantly improved their performance after joining SASSL.

\begin{figure}[ht]
	\centering
	\includegraphics[width=\linewidth,scale=1.00]{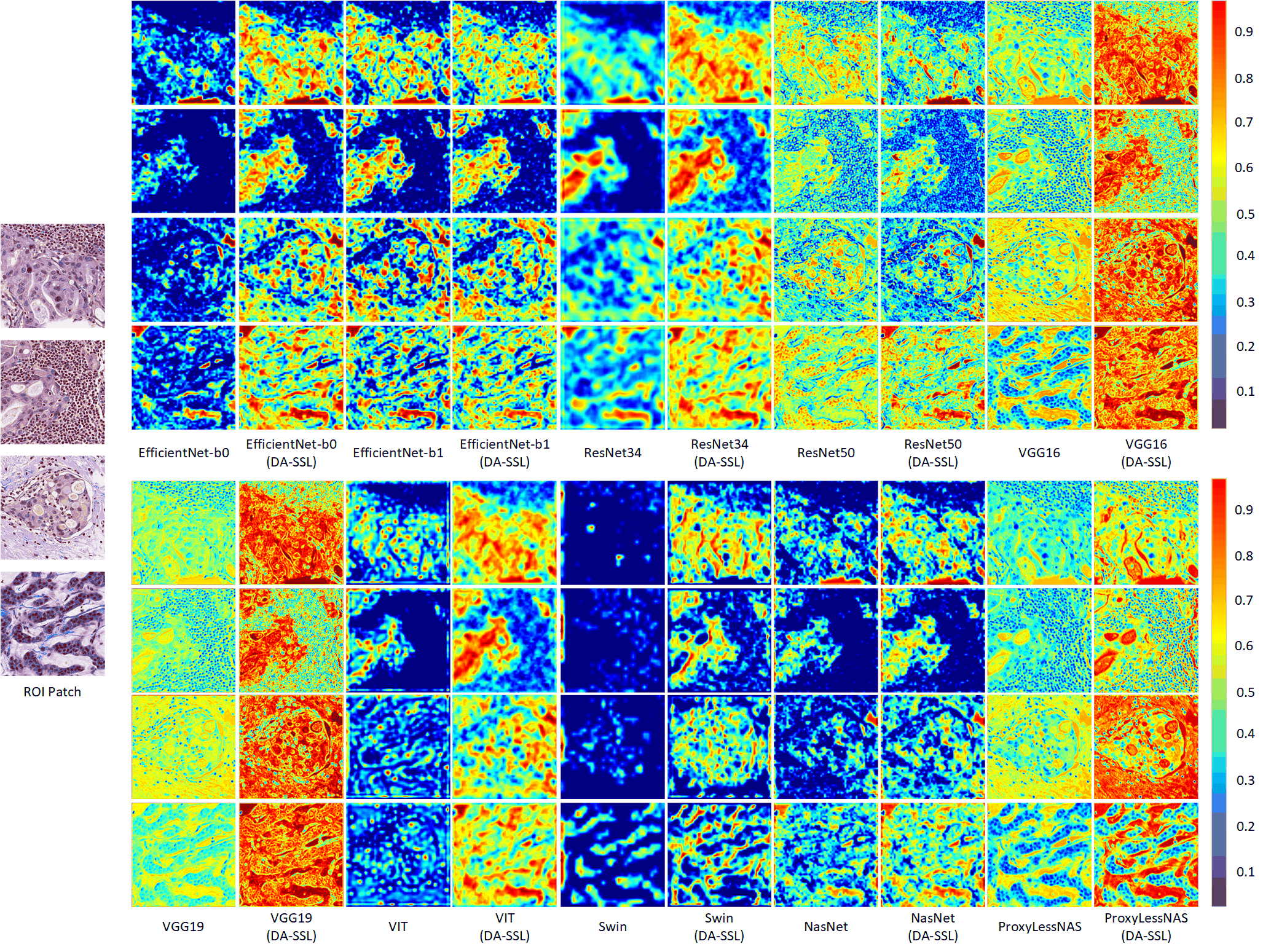}
	\caption{BreastPathQ DataSet HeatMap Changes in Addition of SASSL to Different Backbone Networks.}
	\label{Fig.6}
\end{figure}

\textit{Regression tasks}. Since the downstream model structure of the classification task and the regression task is very similar, we still use the  EfficientNet(\cite{TanLe2019}), ResNet(\cite{HeZhang2015}), RepVGG(\cite{DingZhang2021}), VIT(\cite{DosovitskiyBeyer2021}), Swin(\cite{LiuLin2021}), NASNet(\cite{ZophVasudevan2018}), and ProxylessNAS(\cite{CaiZhu2019}) model as the SOTA benchmark, and sample the same comparison form as the classification task. Compare the original SimSaim with the results of adding SASSL and residual learning. The experimental results are shown in Table 6. From the results, we can see that, similar to the classification task, and each SOTA model has improved steadily after the introduction of SASSL. Among them, the improvement of VIT(\cite{DosovitskiyBeyer2021}) and Swin(\cite{LiuLin2021}) is still very significant. R2 increased by 11\% and 14\%, respectively. Other benchmark models were similar to the classification task, with R2 improvements of 2\%-8\%, respectively. One of the main reasons why the regression task is more effective than the classification task is that the sample size of the PANDA data set is much larger than that of the BreastPathQ data set. This allows the classification model to learn features well even when the initial weights are not good. In the regression task, the model lacks enough samples for feature extraction due to the small number of data set samples. Therefore, the initial weight is significant when the number of samples is small, which highlights the improvement of the model performance by SASSL. In order to better illustrate the feature learning of the regression task, we visualized the initial heating value changes of different SOTA models after SASSL was introduced, as shown in Fig.6. After SASSL is added, the difference in calorific value between the different tissue areas of the model is more obvious, which can help the model locate the lesion and key areas more quickly.

\begin{figure}[ht]
	\centering
	\includegraphics[width=\linewidth,scale=1.00]{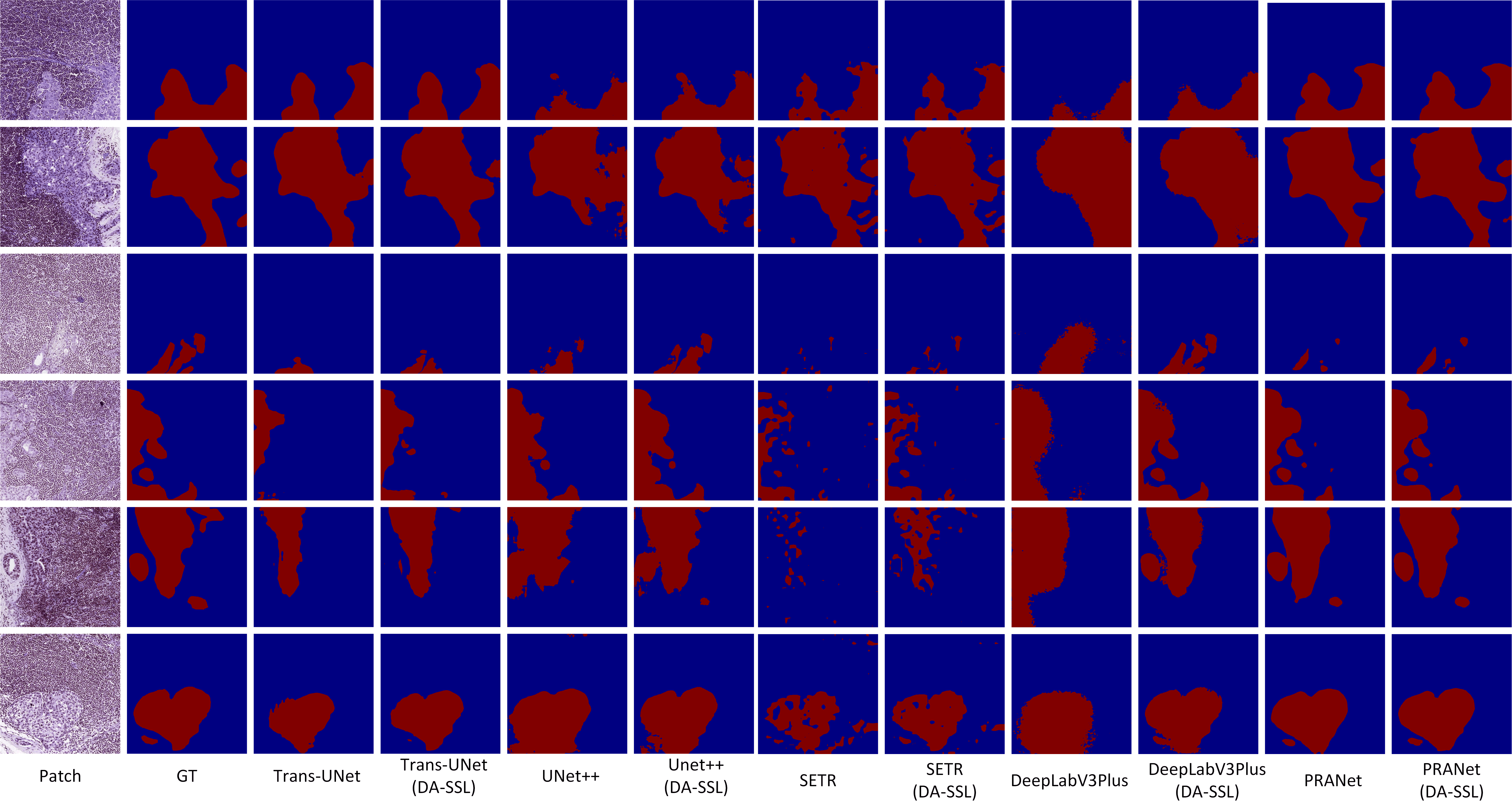}
	\caption{Panda DataSet HeatMap with Differnent Self-Supervisor.}
	\label{Fig.7}
\end{figure}

\begin{figure*}[ht]
	\centering
	\includegraphics[width=0.8\linewidth,scale=1.00]{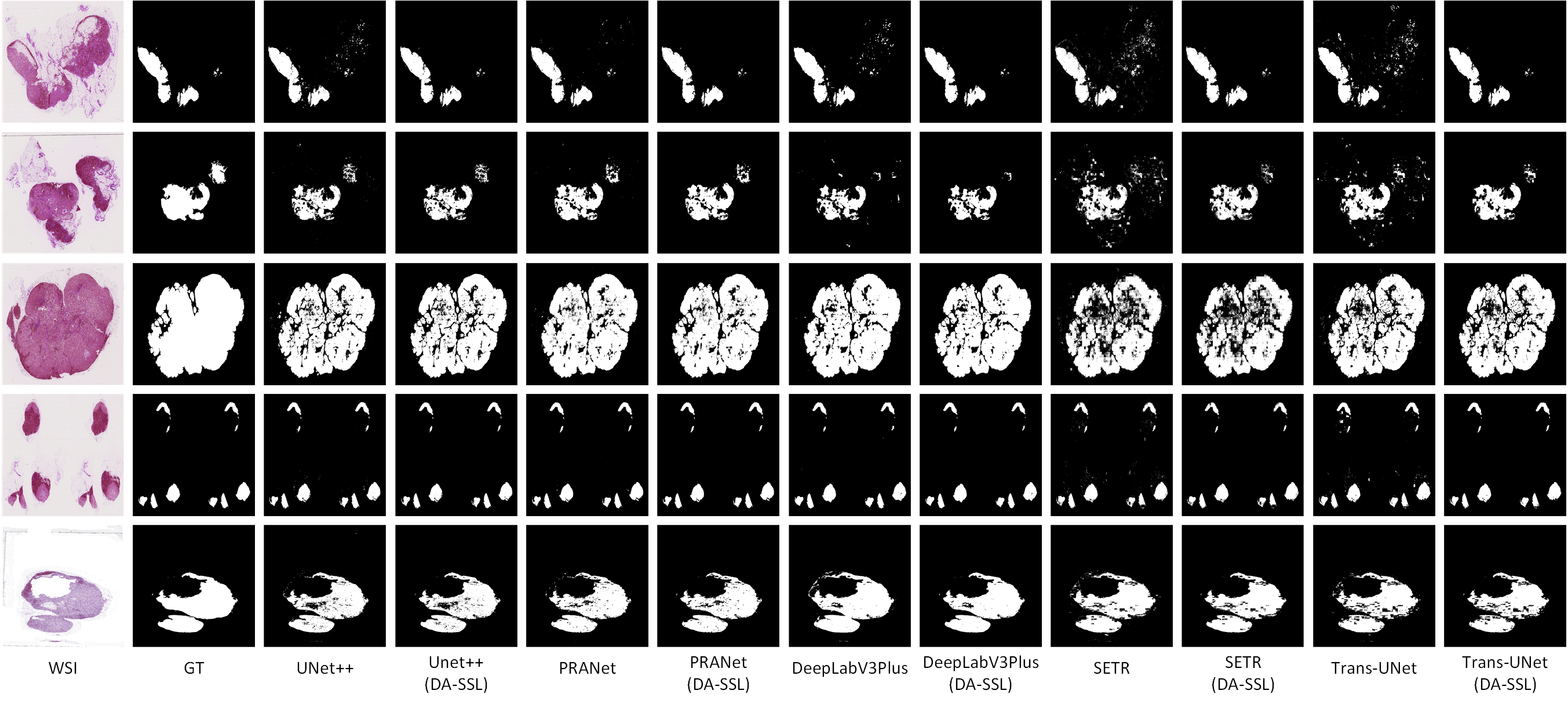}
	\caption{Comparison of each baseline on the CAMELYON16 test set with the pixel-level segmentation results after SASSL is added.}
	\label{Fig.8}
\end{figure*}

\textit{Segmentation tasks}. The segmentation of pathological images is more complicated than classification and regression. For the current general segmentation benchmark model, we have selected UNet$++$(\cite{ZhouSiddiquee2018}), DeepLabV3Plus(\cite{ChenZhu2018}), PRANet(\cite{FanJi2020}), SETR(\cite{ZhengLu2021}), and TransUNet(\cite{ChenLu2021}). Among them, UNet$++$ and DeepLabV3Plus are the general baselines of pathological images, which have been widely used in previous works of pathological image analysis. SETR and TransUNet are relatively novel segmentation models based on the Transformer structure. PRANet is the current SOTA model for segmenting polyps and foot ulcers. It has achieved good results with efficient attention feature modules. The experimental results are shown in Table 7. It can be seen that the performance of the above segmentation model has been significantly improved after adding SASSL. For SETR(\cite{ZhengLu2021}) and TransUNet(\cite{ChenLu2021}) based on the Transformer structure, the performance of the model is significantly improved after SASSL is added, and the Dice coefficient is increased by 7\% and 5\%, respectively. The Dice coefficient of DeepLabV3Plus has also increased a lot, an increase of 6.6\%. The Dice coefficient of other models also increased by 1\%-4\%. In order to better analyze the improvement direction of the model for the segmentation structure, we visualized the segmentation results of each model for the patch, as shown in Fig.7. SASSL can significantly improve the discriminative of the segmentation model for different regions in the image, and this kind of improvement is robust. It will not be subject to large fluctuations caused by color differences. The visualized results of the overall segmentation results of WSIs are shown in Fig.8. From the perspective of the overall segmentation effect, after the introduction of SASSL, the model has significantly improved the segmentation effect of difficult-to-segment regions, reducing the probability of many such regions being segmented as false positives. Furthermore, the edges of the segmentation result are smoother, effectively reducing the appearance of jagged edges.

\begin{table}[ht]
	\centering
	\caption{\newline Comparison of the structural ablation of different segmentater baseline in the CAMELYON16 data set.}
	\begin{tabular}{c|c|c|c}
		\toprule[1pt]
		Method & $PA$ & $Dice$ & $MIoU$ \\
		\toprule[1pt]
		UNet$++$ & 0.905 & 0.840 & 0.766  \\
		UNet$++$(SASSL) & 0.929 & 0.906 & 0.839 \\ 
		\hline
		DeepLabV3Plus & 0.876 & 0.883 & 0.791 \\
		DeepLabV3Plus(SASSL) & 0.905 & 0.923 & 0.857 \\ 
		\hline
		PRANet & 0.935 & 0.933 & 0.867 \\
		PRANet(SASSL) & \textbf{0.948} & \textbf{0.946}& \textbf{0.898} \\ 
		\toprule[1pt]
		SETR & 0.798 & 0.829 & 0.784 \\
		SETR(SASSL) & 0.850 & 0.894 & 0.809 \\ 
		\hline
		TransUNet & 0.813 & 0.843 & 0.808 \\
		TransUNet(SASSL) & \textbf{0.883} & \textbf{0.899} &\textbf{0.818} \\ 
		\toprule[1pt]
	\end{tabular}
	\label{Table7}
\end{table}

\subsection{Conclusion}
Computer-aided diagnosis of histopathological images can improve accuracy while reducing the burden on pathologists. This research proposes a stain-adaptive self-supervised learning framework(SASSL) method. Under un-supervised conditions, stain-adaptive self-supervised learning enables the model to adapt to staining while extracting potential invariance features of WSIs. SASSL automatically mines the potential features of pathological image data from multiple data sets, improves the model's ability to extract deep features, and proves the effectiveness of SASSL on various types of downstream tasks. More importantly, SASSL training methods and pre-training weights can be extended to other research work for histopathological image analysis.

Although the performance of each model has been steadily improved after the introduction of SASSL, the model with the Transformer structure still has the potential for further improvement. By referring to Transformer's self-supervised learning work in the NLP direction, the model based on Transformer structure can be generalized in pathological image analysis problems.

\section*{Acknowledgments}
Acknowledgments should be inserted at the end of the paper, before the
references, not as a footnote to the title. Use the unnumbered
Acknowledgements Head style for the Acknowledgments heading.

\bibliographystyle{model2-names.bst}\biboptions{authoryear}
\bibliography{SASSL}

\begin{thebibliography}{79}
\expandafter\ifx\csname natexlab\endcsname\relax\def\natexlab#1{#1}\fi
\providecommand{\url}[1]{\texttt{#1}}
\providecommand{\href}[2]{#2}
\providecommand{\path}[1]{#1}
\providecommand{\DOIprefix}{doi:}
\providecommand{\ArXivprefix}{arXiv:}
\providecommand{\URLprefix}{URL: }
\providecommand{\Pubmedprefix}{pmid:}
\providecommand{\doi}[1]{\href{http://dx.doi.org/#1}{\path{#1}}}
\providecommand{\Pubmed}[1]{\href{pmid:#1}{\path{#1}}}
\providecommand{\bibinfo}[2]{#2}
\ifx\xfnm\relax \def\xfnm[#1]{\unskip,\space#1}\fi
\bibitem[{Bejnordi et~al.(2016)Bejnordi, Litjens, Timofeeva,
  Otte{-}H{\"{o}}ller, Homeyer, Karssemeijer and van~der
  Laak}]{BejnordiLitjens2016}
\bibinfo{author}{Bejnordi, B.E.}, \bibinfo{author}{Litjens, G.},
  \bibinfo{author}{Timofeeva, N.}, \bibinfo{author}{Otte{-}H{\"{o}}ller, I.},
  \bibinfo{author}{Homeyer, A.}, \bibinfo{author}{Karssemeijer, N.},
  \bibinfo{author}{van~der Laak, J.A.W.M.}, \bibinfo{year}{2016}.
\newblock \bibinfo{title}{Stain specific standardization of whole-slide
  histopathological images}.
\newblock \bibinfo{journal}{IEEE Trans Med Imaging} \bibinfo{volume}{35(2)},
  \bibinfo{pages}{404--415}.
\newblock \DOIprefix\doi{10.1109/TMI.2015.2476509}.
\bibitem[{BenTaieb and Hamarneh(2018)}]{BenTaiebHamarneh2018}
\bibinfo{author}{BenTaieb, A.}, \bibinfo{author}{Hamarneh, G.},
  \bibinfo{year}{2018}.
\newblock \bibinfo{title}{Adversarial stain transfer for histopathology image
  analysis}.
\newblock \bibinfo{journal}{IEEE Trans Med Imaging} \bibinfo{volume}{37(3)},
  \bibinfo{pages}{792--802}.
\newblock \DOIprefix\doi{10.1109/TMI.2017.2781228}.
\bibitem[{Bándi et~al.(2019)Bándi, Geessink, Manson, Van~Dijk, Balkenhol,
  Hermsen, Ehteshami~Bejnordi, Lee, Paeng, Zhong, Li, Zanjani, Zinger, Fukuta,
  Komura, Ovtcharov, Cheng, Zeng, Thagaard, Dahl, Lin, Chen, Jacobsson,
  Hedlund, Çetin, Halıcı, Jackson, Chen, Both, Franke, Küsters-Vandevelde,
  Vreuls, Bult, van Ginneken, van~der Laak and Litjens}]{PeterOscar2019}
\bibinfo{author}{Bándi, P.}, \bibinfo{author}{Geessink, O.},
  \bibinfo{author}{Manson, Q.}, \bibinfo{author}{Van~Dijk, M.},
  \bibinfo{author}{Balkenhol, M.}, \bibinfo{author}{Hermsen, M.},
  \bibinfo{author}{Ehteshami~Bejnordi, B.}, \bibinfo{author}{Lee, B.},
  \bibinfo{author}{Paeng, K.}, \bibinfo{author}{Zhong, A.},
  \bibinfo{author}{Li, Q.}, \bibinfo{author}{Zanjani, F.G.},
  \bibinfo{author}{Zinger, S.}, \bibinfo{author}{Fukuta, K.},
  \bibinfo{author}{Komura, D.}, \bibinfo{author}{Ovtcharov, V.},
  \bibinfo{author}{Cheng, S.}, \bibinfo{author}{Zeng, S.},
  \bibinfo{author}{Thagaard, J.}, \bibinfo{author}{Dahl, A.B.},
  \bibinfo{author}{Lin, H.}, \bibinfo{author}{Chen, H.},
  \bibinfo{author}{Jacobsson, L.}, \bibinfo{author}{Hedlund, M.},
  \bibinfo{author}{Çetin, M.}, \bibinfo{author}{Halıcı, E.},
  \bibinfo{author}{Jackson, H.}, \bibinfo{author}{Chen, R.},
  \bibinfo{author}{Both, F.}, \bibinfo{author}{Franke, J.},
  \bibinfo{author}{Küsters-Vandevelde, H.}, \bibinfo{author}{Vreuls, W.},
  \bibinfo{author}{Bult, P.}, \bibinfo{author}{van Ginneken, B.},
  \bibinfo{author}{van~der Laak, J.}, \bibinfo{author}{Litjens, G.},
  \bibinfo{year}{2019}.
\newblock \bibinfo{title}{From detection of individual metastases to
  classification of lymph node status at the patient level: The camelyon17
  challenge}.
\newblock \bibinfo{journal}{IEEE Trans Med Imaging} \bibinfo{volume}{38(2)},
  \bibinfo{pages}{550--560}.
\newblock \DOIprefix\doi{10.1109/TMI.2018.2867350}.
\bibitem[{Cai et~al.(2019)Cai, Zhu and Han}]{CaiZhu2019}
\bibinfo{author}{Cai, H.}, \bibinfo{author}{Zhu, L.}, \bibinfo{author}{Han,
  S.}, \bibinfo{year}{2019}.
\newblock \bibinfo{title}{Proxylessnas: Direct neural architecture search on
  target task and hardware}, in: \bibinfo{booktitle}{International Conference
  on Learning Representations (ICLR)}.
\newblock \DOIprefix\doi{10.48550/arXiv.1812.00332}.
\bibitem[{Campanella et~al.(2019)Campanella, Hanna, Geneslaw, Miraflor and
  Fuchs}]{CampanellaHanna2019}
\bibinfo{author}{Campanella, G.}, \bibinfo{author}{Hanna, M.G.},
  \bibinfo{author}{Geneslaw, L.}, \bibinfo{author}{Miraflor, A.},
  \bibinfo{author}{Fuchs, T.J.}, \bibinfo{year}{2019}.
\newblock \bibinfo{title}{Clinical-grade computational pathology using weakly
  supervised deep learning on whole slide images}.
\newblock \bibinfo{journal}{Nature Medicine} \bibinfo{volume}{25(8)},
  \bibinfo{pages}{1301–1309}.
\newblock \DOIprefix\doi{10.1038/s41591-019-0508-1}.
\bibitem[{Chen et~al.(2016)Chen, Wang and Heng}]{ChenWang2016}
\bibinfo{author}{Chen, H.}, \bibinfo{author}{Wang, X.}, \bibinfo{author}{Heng,
  P.}, \bibinfo{year}{2016}.
\newblock \bibinfo{title}{Automated mitosis detection with deep regression
  networks}, in: \bibinfo{booktitle}{IEEE International Symposium on Biomedical
  Imaging (ISBI)}, pp. \bibinfo{pages}{1204--1207}.
\newblock \DOIprefix\doi{10.1109/ISBI.2016.7493482}.
\bibitem[{Chen et~al.(2021)Chen, Lu, Yu, Luo, Adeli, Wang, Lu, Yuille and
  Zhou}]{ChenLu2021}
\bibinfo{author}{Chen, J.}, \bibinfo{author}{Lu, Y.}, \bibinfo{author}{Yu, Q.},
  \bibinfo{author}{Luo, X.}, \bibinfo{author}{Adeli, E.},
  \bibinfo{author}{Wang, Y.}, \bibinfo{author}{Lu, L.},
  \bibinfo{author}{Yuille, A.L.}, \bibinfo{author}{Zhou, Y.},
  \bibinfo{year}{2021}.
\newblock \bibinfo{title}{Transunet: Transformers make strong encoders for
  medical image segmentation}.
\newblock \bibinfo{journal}{arXiv} \DOIprefix\doi{10.48550/arXiv.2102.04306}.
\bibitem[{Chen et~al.(2018)Chen, Zhu, Papandreou, Schroff and
  Adam}]{ChenZhu2018}
\bibinfo{author}{Chen, L.}, \bibinfo{author}{Zhu, Y.},
  \bibinfo{author}{Papandreou, G.}, \bibinfo{author}{Schroff, F.},
  \bibinfo{author}{Adam, H.}, \bibinfo{year}{2018}.
\newblock \bibinfo{title}{Encoder-decoder with atrous separable convolution for
  semantic image segmentation}, in: \bibinfo{booktitle}{European Conference
  Computer Vision (ECCV)}, pp. \bibinfo{pages}{833--851}.
\newblock \DOIprefix\doi{10.1007/978-3-030-01234-2\_49}.
\bibitem[{Chen et~al.(2020a)Chen, Kornblith, Norouzi and
  Hinton}]{ChenKornblith2020v1}
\bibinfo{author}{Chen, T.}, \bibinfo{author}{Kornblith, S.},
  \bibinfo{author}{Norouzi, M.}, \bibinfo{author}{Hinton, G.E.},
  \bibinfo{year}{2020}a.
\newblock \bibinfo{title}{A simple framework for contrastive learning of visual
  representations}, in: \bibinfo{booktitle}{International Conference on Machine
  Learning (ICML)}, pp. \bibinfo{pages}{1597--1607}.
\newblock \DOIprefix\doi{10.48550/arXiv.2002.05709}.
\bibitem[{Chen et~al.(2020b)Chen, Kornblith, Swersky, Norouzi and
  Hinton}]{ChenKornblith2020v2}
\bibinfo{author}{Chen, T.}, \bibinfo{author}{Kornblith, S.},
  \bibinfo{author}{Swersky, K.}, \bibinfo{author}{Norouzi, M.},
  \bibinfo{author}{Hinton, G.E.}, \bibinfo{year}{2020}b.
\newblock \bibinfo{title}{Big self-supervised models are strong semi-supervised
  learners}, in: \bibinfo{booktitle}{Neural Information Processing Systems
  (NIPS)}.
\newblock \DOIprefix\doi{10.48550/arXiv.2006.10029}.
\bibitem[{Chen et~al.(2020c)Chen, Fan, Girshick and He}]{ChenFan2020}
\bibinfo{author}{Chen, X.}, \bibinfo{author}{Fan, H.},
  \bibinfo{author}{Girshick, R.B.}, \bibinfo{author}{He, K.},
  \bibinfo{year}{2020}c.
\newblock \bibinfo{title}{Improved baselines with momentum contrastive
  learning}, in: \bibinfo{booktitle}{arXiv}.
\newblock \DOIprefix\doi{10.48550/arXiv.2003.04297}.
\bibitem[{Chen and He(2021)}]{ChenHe2021}
\bibinfo{author}{Chen, X.}, \bibinfo{author}{He, K.}, \bibinfo{year}{2021}.
\newblock \bibinfo{title}{Exploring simple siamese representation learning},
  in: \bibinfo{booktitle}{IEEE Computer Vision and Pattern Recognition (CVPR)},
  pp. \bibinfo{pages}{15750--15758}.
\newblock \DOIprefix\doi{10.1109/CVPR46437.2021.01549}.
\bibitem[{{\c{C}}i{\c{c}}ek et~al.(2016){\c{C}}i{\c{c}}ek, Abdulkadir,
  Lienkamp, Brox and Ronneberger}]{AbdulkadirLienkamp2016}
\bibinfo{author}{{\c{C}}i{\c{c}}ek, {\"{O}}.}, \bibinfo{author}{Abdulkadir,
  A.}, \bibinfo{author}{Lienkamp, S.S.}, \bibinfo{author}{Brox, T.},
  \bibinfo{author}{Ronneberger, O.}, \bibinfo{year}{2016}.
\newblock \bibinfo{title}{3d u-net: Learning dense volumetric segmentation from
  sparse annotation}, in: \bibinfo{editor}{Ourselin, S.},
  \bibinfo{editor}{Joskowicz, L.}, \bibinfo{editor}{Sabuncu, M.R.},
  \bibinfo{editor}{{\"{U}}nal, G.B.}, \bibinfo{editor}{Wells, W.} (Eds.),
  \bibinfo{booktitle}{Medical Image Computing and Computer-Assisted
  Intervention (MICCAI)}, pp. \bibinfo{pages}{424--432}.
\newblock \DOIprefix\doi{10.1007/978-3-319-46723-8\_49}.
\bibitem[{Ding et~al.(2021)Ding, Zhang, Ma, Han, Ding and Sun}]{DingZhang2021}
\bibinfo{author}{Ding, X.}, \bibinfo{author}{Zhang, X.}, \bibinfo{author}{Ma,
  N.}, \bibinfo{author}{Han, J.}, \bibinfo{author}{Ding, G.},
  \bibinfo{author}{Sun, J.}, \bibinfo{year}{2021}.
\newblock \bibinfo{title}{Repvgg: Making vgg-style convnets great again}, in:
  \bibinfo{booktitle}{IEEE Computer Vision and Pattern Recognition (CVPR)}, pp.
  \bibinfo{pages}{13733--13742}.
\newblock \DOIprefix\doi{10.1109/CVPR46437.2021.01352}.
\bibitem[{Dosovitskiy et~al.(2021)Dosovitskiy, Beyer, Kolesnikov, Weissenborn,
  Zhai, Unterthiner, Dehghani, Minderer, Heigold, Gelly, Uszkoreit and
  Houlsby}]{DosovitskiyBeyer2021}
\bibinfo{author}{Dosovitskiy, A.}, \bibinfo{author}{Beyer, L.},
  \bibinfo{author}{Kolesnikov, A.}, \bibinfo{author}{Weissenborn, D.},
  \bibinfo{author}{Zhai, X.}, \bibinfo{author}{Unterthiner, T.},
  \bibinfo{author}{Dehghani, M.}, \bibinfo{author}{Minderer, M.},
  \bibinfo{author}{Heigold, G.}, \bibinfo{author}{Gelly, S.},
  \bibinfo{author}{Uszkoreit, J.}, \bibinfo{author}{Houlsby, N.},
  \bibinfo{year}{2021}.
\newblock \bibinfo{title}{An image is worth 16x16 words: Transformers for image
  recognition at scale}, in: \bibinfo{booktitle}{International Conference on
  Learning Representations(ICLR)}.
\newblock \DOIprefix\doi{10.48550/arXiv.2010.11929}.
\bibitem[{Ehteshami~Bejnordi et~al.(2017)Ehteshami~Bejnordi, Veta, Johannes~van
  Diest, van Ginneken, Karssemeijer, Litjens and van~der Laak}]{BabakMitko2017}
\bibinfo{author}{Ehteshami~Bejnordi, B.}, \bibinfo{author}{Veta, M.},
  \bibinfo{author}{Johannes~van Diest, P.}, \bibinfo{author}{van Ginneken, B.},
  \bibinfo{author}{Karssemeijer, N.}, \bibinfo{author}{Litjens, G.},
  \bibinfo{author}{van~der Laak, J.A.W.M.}, \bibinfo{year}{2017}.
\newblock \bibinfo{title}{{Diagnostic Assessment of Deep Learning Algorithms
  for Detection of Lymph Node Metastases in Women With Breast Cancer}}.
\newblock \bibinfo{journal}{JAMA} \bibinfo{volume}{318(22)},
  \bibinfo{pages}{2199--2210}.
\newblock \DOIprefix\doi{10.1001/jama.2017.14585}.
\bibitem[{Emad et~al.(2008)Emad, Maysa, Andrew, Christopher and
  Ian}]{EmadMaysa2008}
\bibinfo{author}{Emad, A.}, \bibinfo{author}{Maysa, E.R.},
  \bibinfo{author}{Andrew, H.S.}, \bibinfo{author}{Christopher, W.E.},
  \bibinfo{author}{Ian, O.E.}, \bibinfo{year}{2008}.
\newblock \bibinfo{title}{Prognostic significance of nottingham histologic
  grade in invasive breast carcinoma}.
\newblock \bibinfo{journal}{American Society of Clinical Oncology}
  \bibinfo{volume}{26(19)}, \bibinfo{pages}{3153–3158}.
\newblock \DOIprefix\doi{10.1200/JCO.2007.15.5986}.
\bibitem[{Epstein et~al.(2006)Epstein, Allsbrook, Amin and
  Egevad}]{Bjartell2005}
\bibinfo{author}{Epstein, J.I.}, \bibinfo{author}{Allsbrook, W.~C., J.},
  \bibinfo{author}{Amin, M.B.}, \bibinfo{author}{Egevad, L.L.},
  \bibinfo{year}{2006}.
\newblock \bibinfo{title}{The 2005 international society of urological
  pathology (isup) consensus conference on gleason grading of prostatic
  carcinoma}.
\newblock \bibinfo{journal}{European Urology} \bibinfo{volume}{49(4)},
  \bibinfo{pages}{758--759}.
\newblock \DOIprefix\doi{0.1097/01.pas.0000173646.99337.b1}.
\bibitem[{Fabius et~al.(2015)Fabius, van Amersfoort and
  Kingma}]{FabiusAmersfoort2015}
\bibinfo{author}{Fabius, O.}, \bibinfo{author}{van Amersfoort, J.R.},
  \bibinfo{author}{Kingma, D.P.}, \bibinfo{year}{2015}.
\newblock \bibinfo{title}{Variational recurrent auto-encoders}, in:
  \bibinfo{booktitle}{International Conference on Learning Representations
  (ICLR)}.
\newblock \DOIprefix\doi{10.48550/arXiv.1412.6581}.
\bibitem[{Fan et~al.(2020)Fan, Ji, Zhou, Chen, Fu, Shen and Shao}]{FanJi2020}
\bibinfo{author}{Fan, D.}, \bibinfo{author}{Ji, G.}, \bibinfo{author}{Zhou,
  T.}, \bibinfo{author}{Chen, G.}, \bibinfo{author}{Fu, H.},
  \bibinfo{author}{Shen, J.}, \bibinfo{author}{Shao, L.}, \bibinfo{year}{2020}.
\newblock \bibinfo{title}{Pranet: Parallel reverse attention network for polyp
  segmentation}, in: \bibinfo{booktitle}{Medical Image Computing and Computer
  Assisted Intervention (MICCAI)}, pp. \bibinfo{pages}{263--273}.
\newblock \DOIprefix\doi{10.1007/978-3-030-59725-2\_26}.
\bibitem[{Gao et~al.(2021)Gao, Cheng, Zhao, Zhang, Yang and
  Torr}]{GaoCheng2021}
\bibinfo{author}{Gao, S.H.}, \bibinfo{author}{Cheng, M.M.},
  \bibinfo{author}{Zhao, K.}, \bibinfo{author}{Zhang, X.Y.},
  \bibinfo{author}{Yang, M.H.}, \bibinfo{author}{Torr, P.},
  \bibinfo{year}{2021}.
\newblock \bibinfo{title}{Res2net: A new multi-scale backbone architecture}.
\newblock \bibinfo{journal}{IEEE Transactions on Pattern Analysis and Machine
  Intelligence(T-PAMI)} \bibinfo{volume}{43(2)}, \bibinfo{pages}{652--662}.
\newblock \DOIprefix\doi{10.1109/TPAMI.2019.2938758}.
\bibitem[{Gao et~al.(2017)Gao, Wang, Zhou and Zhang}]{GaoWang2017}
\bibinfo{author}{Gao, Z.}, \bibinfo{author}{Wang, L.}, \bibinfo{author}{Zhou,
  L.}, \bibinfo{author}{Zhang, J.}, \bibinfo{year}{2017}.
\newblock \bibinfo{title}{Hep-2 cell image classification with deep
  convolutional neural networks}.
\newblock \bibinfo{journal}{IEEE J. Biomed. Health Informatics}
  \bibinfo{volume}{21(2)}, \bibinfo{pages}{416--428}.
\newblock \DOIprefix\doi{10.1109/JBHI.2016.2526603}.
\bibitem[{Geert et~al.(2018)Geert, Peter, Babak, Oscar, Maschenka, Peter,
  Altuna, Meyke, van~de Loo~Rob and Rob}]{GeertPeter2018}
\bibinfo{author}{Geert, L.}, \bibinfo{author}{Peter, B.},
  \bibinfo{author}{Babak, E.B.}, \bibinfo{author}{Oscar, G.},
  \bibinfo{author}{Maschenka, B.}, \bibinfo{author}{Peter, B.},
  \bibinfo{author}{Altuna, H.}, \bibinfo{author}{Meyke, H.},
  \bibinfo{author}{van~de Loo~Rob}, \bibinfo{author}{Rob, V.},
  \bibinfo{year}{2018}.
\newblock \bibinfo{title}{1399 he-stained sentinel lymph node sections of
  breast cancer patients: the camelyon dataset}.
\newblock \bibinfo{journal}{GigaScience} \bibinfo{volume}{7(6)},
  \bibinfo{pages}{416--428}.
\newblock \DOIprefix\doi{10.1093/gigascience/giy065}.
\bibitem[{Graham et~al.(2019)Graham, Chen, Gamper, Dou, Heng, Snead, Tsang and
  Rajpoot}]{GrahamChen2019}
\bibinfo{author}{Graham, S.}, \bibinfo{author}{Chen, H.},
  \bibinfo{author}{Gamper, J.}, \bibinfo{author}{Dou, Q.},
  \bibinfo{author}{Heng, P.}, \bibinfo{author}{Snead, D.R.J.},
  \bibinfo{author}{Tsang, Y.}, \bibinfo{author}{Rajpoot, N.M.},
  \bibinfo{year}{2019}.
\newblock \bibinfo{title}{Mild-net: Minimal information loss dilated network
  for gland instance segmentation in colon histology images}.
\newblock \bibinfo{journal}{Med Image Anal} \bibinfo{volume}{52},
  \bibinfo{pages}{199--211}.
\newblock \DOIprefix\doi{10.1016/j.media.2018.12.001}.
\bibitem[{Grill et~al.(2020)Grill, Strub, Altch{\'{e}}, Tallec, Richemond,
  Buchatskaya, Doersch, Pires, Guo, Azar, Piot, Kavukcuoglu, Munos and
  Valko}]{GrillStrub2020}
\bibinfo{author}{Grill, J.}, \bibinfo{author}{Strub, F.},
  \bibinfo{author}{Altch{\'{e}}, F.}, \bibinfo{author}{Tallec, C.},
  \bibinfo{author}{Richemond, P.H.}, \bibinfo{author}{Buchatskaya, E.},
  \bibinfo{author}{Doersch, C.}, \bibinfo{author}{Pires, B.{\'{A}}.},
  \bibinfo{author}{Guo, Z.}, \bibinfo{author}{Azar, M.G.},
  \bibinfo{author}{Piot, B.}, \bibinfo{author}{Kavukcuoglu, K.},
  \bibinfo{author}{Munos, R.}, \bibinfo{author}{Valko, M.},
  \bibinfo{year}{2020}.
\newblock \bibinfo{title}{Bootstrap your own latent - {A} new approach to
  self-supervised learning}, in: \bibinfo{booktitle}{Neural Information
  Processing Systems (NIPS)}.
\newblock \DOIprefix\doi{10.48550/arXiv.2006.07733}.
\bibitem[{Gurcan et~al.(2009)Gurcan, Boucheron, Can, Madabhushi, Rajpoot and
  Yener}]{GurcanBoucheron2009}
\bibinfo{author}{Gurcan, M.N.}, \bibinfo{author}{Boucheron, L.E.},
  \bibinfo{author}{Can, A.}, \bibinfo{author}{Madabhushi, A.},
  \bibinfo{author}{Rajpoot, N.M.}, \bibinfo{author}{Yener, B.},
  \bibinfo{year}{2009}.
\newblock \bibinfo{title}{Histopathological image analysis: A review}.
\newblock \bibinfo{journal}{IEEE Reviews in Biomedical Engineering}
  \bibinfo{volume}{2}, \bibinfo{pages}{147--171}.
\newblock \DOIprefix\doi{10.1109/RBME.2009.2034865}.
\bibitem[{Gutmann and Hyv{\"{a}}rinen(2012)}]{Gutmannrinen2012}
\bibinfo{author}{Gutmann, M.}, \bibinfo{author}{Hyv{\"{a}}rinen, A.},
  \bibinfo{year}{2012}.
\newblock \bibinfo{title}{Noise-contrastive estimation of unnormalized
  statistical models, with applications to natural image statistics}.
\newblock \bibinfo{journal}{Machine Learning Research} \bibinfo{volume}{13},
  \bibinfo{pages}{307--361}.
\newblock \DOIprefix\doi{10.5555/2503308.2188396}.
\bibitem[{He et~al.(2020)He, Fan, Wu, Xie and Girshick}]{HeFan2020}
\bibinfo{author}{He, K.}, \bibinfo{author}{Fan, H.}, \bibinfo{author}{Wu, Y.},
  \bibinfo{author}{Xie, S.}, \bibinfo{author}{Girshick, R.B.},
  \bibinfo{year}{2020}.
\newblock \bibinfo{title}{Momentum contrast for unsupervised visual
  representation learning}, in: \bibinfo{booktitle}{IEEE Computer Vision and
  Pattern Recognition (CVPR)}, pp. \bibinfo{pages}{9726--9735}.
\newblock \DOIprefix\doi{10.1109/CVPR42600.2020.00975}.
\bibitem[{He et~al.(2016)He, Zhang, Ren and Sun}]{HeZhang2015}
\bibinfo{author}{He, K.}, \bibinfo{author}{Zhang, X.}, \bibinfo{author}{Ren,
  S.}, \bibinfo{author}{Sun, J.}, \bibinfo{year}{2016}.
\newblock \bibinfo{title}{Deep residual learning for image recognition}, in:
  \bibinfo{booktitle}{IEEE Computer Vision and Pattern Recognition (CVPR)}, pp.
  \bibinfo{pages}{770--778}.
\newblock \DOIprefix\doi{10.1109/CVPR.2016.90}.
\bibitem[{Hou et~al.(2021)Hou, Zhou and Feng}]{HouZhou2021}
\bibinfo{author}{Hou, Q.}, \bibinfo{author}{Zhou, D.}, \bibinfo{author}{Feng,
  J.}, \bibinfo{year}{2021}.
\newblock \bibinfo{title}{Coordinate attention for efficient mobile network
  design}, in: \bibinfo{booktitle}{IEEE Computer Vision and Pattern Recognition
  (CVPR)}, pp. \bibinfo{pages}{13713--13722}.
\newblock \DOIprefix\doi{10.1109/CVPR46437.2021.01350}.
\bibitem[{Hu et~al.(2019)Hu, Tang, Chang, Fan, Lai and Xu}]{HuTang2019}
\bibinfo{author}{Hu, B.}, \bibinfo{author}{Tang, Y.}, \bibinfo{author}{Chang,
  E.I.}, \bibinfo{author}{Fan, Y.}, \bibinfo{author}{Lai, M.},
  \bibinfo{author}{Xu, Y.}, \bibinfo{year}{2019}.
\newblock \bibinfo{title}{Unsupervised learning for cell-level visual
  representation in histopathology images with generative adversarial
  networks}.
\newblock \bibinfo{journal}{IEEE J. Biomed. Health Informatics}
  \bibinfo{volume}{23(3)}, \bibinfo{pages}{1316--1328}.
\newblock \DOIprefix\doi{10.1109/JBHI.2018.2852639}.
\bibitem[{Hu et~al.(2018)Hu, Shen and Sun}]{HuShen2018}
\bibinfo{author}{Hu, J.}, \bibinfo{author}{Shen, L.}, \bibinfo{author}{Sun,
  G.}, \bibinfo{year}{2018}.
\newblock \bibinfo{title}{Squeeze-and-excitation networks}, in:
  \bibinfo{booktitle}{IEEE Computer Vision and Pattern Recognition (CVPR)}, pp.
  \bibinfo{pages}{7132--7141}.
\newblock \DOIprefix\doi{10.1109/CVPR.2018.00745}.
\bibitem[{Huang et~al.(2017)Huang, Liu, Van Der~Maaten and
  Weinberger}]{HuangLiu2017}
\bibinfo{author}{Huang, G.}, \bibinfo{author}{Liu, Z.}, \bibinfo{author}{Van
  Der~Maaten, L.}, \bibinfo{author}{Weinberger, K.Q.}, \bibinfo{year}{2017}.
\newblock \bibinfo{title}{Densely connected convolutional networks}, in:
  \bibinfo{booktitle}{IEEE Computer Vision and Pattern Recognition (CVPR)}, pp.
  \bibinfo{pages}{2261--2269}.
\newblock \DOIprefix\doi{10.1109/CVPR.2017.243}.
\bibitem[{Jing and Tian(2021)}]{JingTian2019}
\bibinfo{author}{Jing, L.}, \bibinfo{author}{Tian, Y.}, \bibinfo{year}{2021}.
\newblock \bibinfo{title}{Self-supervised visual feature learning with deep
  neural networks: A survey}.
\newblock \bibinfo{journal}{IEEE Transactions on Pattern Analysis and Machine
  Intelligence (T-PAMI)} \bibinfo{volume}{43(11)}, \bibinfo{pages}{4037--4058}.
\newblock \DOIprefix\doi{10.1109/TPAMI.2020.2992393}.
\bibitem[{Kashif et~al.(2016)Kashif, Raza, Sirinukunwattana, Arif and
  Rajpoot}]{KashifRaza2016}
\bibinfo{author}{Kashif, M.N.}, \bibinfo{author}{Raza, S.},
  \bibinfo{author}{Sirinukunwattana, K.}, \bibinfo{author}{Arif, M.},
  \bibinfo{author}{Rajpoot, N.M.}, \bibinfo{year}{2016}.
\newblock \bibinfo{title}{Handcrafted features with convolutional neural
  networks for detection of tumor cells in histology images}, in:
  \bibinfo{booktitle}{IEEE International Symposium on Biomedical Imaging
  (ISBI)}, pp. \bibinfo{pages}{1029--1032}.
\newblock \DOIprefix\doi{10.1109/ISBI.2016.7493441}.
\bibitem[{Ke et~al.(2021)Ke, Shen, Liang and Shen}]{KeJing}
\bibinfo{author}{Ke, J.}, \bibinfo{author}{Shen, Y.}, \bibinfo{author}{Liang,
  X.}, \bibinfo{author}{Shen, D.}, \bibinfo{year}{2021}.
\newblock \bibinfo{title}{Contrastive learning based stain normalization across
  multiple tumor in histopathology}, in: \bibinfo{booktitle}{Medical Image
  Computing and Computer Assisted Intervention (MICCAI)}, pp.
  \bibinfo{pages}{571--580}.
\newblock \DOIprefix\doi{10.1007/978-3-030-87237-3\_55}.
\bibitem[{Kervadec et~al.(2021)Kervadec, Bouchtiba, Desrosiers, Granger, Dolz
  and Ayed}]{KervadecBouchtiba2021}
\bibinfo{author}{Kervadec, H.}, \bibinfo{author}{Bouchtiba, J.},
  \bibinfo{author}{Desrosiers, C.}, \bibinfo{author}{Granger, E.},
  \bibinfo{author}{Dolz, J.}, \bibinfo{author}{Ayed, I.B.},
  \bibinfo{year}{2021}.
\newblock \bibinfo{title}{Boundary loss for highly unbalanced segmentation}.
\newblock \bibinfo{journal}{Med Image Anal} \bibinfo{volume}{67},
  \bibinfo{pages}{101851--101869}.
\newblock \DOIprefix\doi{10.1016/j.media.2020.101851}.
\bibitem[{Khan et~al.(2014)Khan, Rajpoot, Treanor and Magee}]{Khan}
\bibinfo{author}{Khan, A.M.}, \bibinfo{author}{Rajpoot, N.},
  \bibinfo{author}{Treanor, D.}, \bibinfo{author}{Magee, D.},
  \bibinfo{year}{2014}.
\newblock \bibinfo{title}{A nonlinear mapping approach to stain normalization
  in digital histopathology images using image-specific color deconvolution}.
\newblock \bibinfo{journal}{IEEE Trans Biomedical Engineering}
  \bibinfo{volume}{61(6)}, \bibinfo{pages}{1729--1738}.
\newblock \DOIprefix\doi{10.1109/TBME.2014.2303294}.
\bibitem[{Kiani et~al.(2020)Kiani, Uyumazturk, Rajpurkar, Wang and
  Shen}]{KianiUyumazturk2020}
\bibinfo{author}{Kiani, A.}, \bibinfo{author}{Uyumazturk, B.},
  \bibinfo{author}{Rajpurkar, P.}, \bibinfo{author}{Wang, A.},
  \bibinfo{author}{Shen, J.}, \bibinfo{year}{2020}.
\newblock \bibinfo{title}{Impact of a deep learning assistant on the
  histopathologic classification of liver cancer}.
\newblock \bibinfo{journal}{npj Digital Medicine} \bibinfo{volume}{23(3)},
  \bibinfo{pages}{1--8}.
\newblock \DOIprefix\doi{10.1038/s41746-020-0232-8}.
\bibitem[{Kim et~al.(2021)Kim, Jang, Lee, Park, Min, Hong, Park, Lee, Kim,
  Hong, Jung, Liu, Rajkumar, Khened, Krishnamurthi, Yang, Wang, Han and
  Choi}]{KimJang2021}
\bibinfo{author}{Kim, Y.J.}, \bibinfo{author}{Jang, H.}, \bibinfo{author}{Lee,
  K.}, \bibinfo{author}{Park, S.}, \bibinfo{author}{Min, S.},
  \bibinfo{author}{Hong, C.}, \bibinfo{author}{Park, J.H.},
  \bibinfo{author}{Lee, K.}, \bibinfo{author}{Kim, J.}, \bibinfo{author}{Hong,
  W.}, \bibinfo{author}{Jung, H.}, \bibinfo{author}{Liu, Y.},
  \bibinfo{author}{Rajkumar, H.}, \bibinfo{author}{Khened, M.},
  \bibinfo{author}{Krishnamurthi, G.}, \bibinfo{author}{Yang, S.},
  \bibinfo{author}{Wang, X.}, \bibinfo{author}{Han, C.H.},
  \bibinfo{author}{Choi, J.}, \bibinfo{year}{2021}.
\newblock \bibinfo{title}{{PAIP} 2019: Liver cancer segmentation challenge}.
\newblock \bibinfo{journal}{Med Image Anal} \bibinfo{volume}{67},
  \bibinfo{pages}{101854}.
\newblock \DOIprefix\doi{10.1016/j.media.2020.101854}.
\bibitem[{Kirk et~al.(2016)Kirk, Lee, Sadow, Levine, Roche, Bonaccio and
  Filiippini}]{tcgacoda}
\bibinfo{author}{Kirk, S.}, \bibinfo{author}{Lee, Y.}, \bibinfo{author}{Sadow,
  C.A.}, \bibinfo{author}{Levine, S.}, \bibinfo{author}{Roche, C.},
  \bibinfo{author}{Bonaccio, E.}, \bibinfo{author}{Filiippini, J.},
  \bibinfo{year}{2016}.
\newblock \bibinfo{title}{Radiology data from the cancer genome atlas colon
  adenocarcinoma tcga-coad collection}.
\newblock \bibinfo{journal}{The Cancer Imaging Archive (TCIA) Public Access}
  \bibinfo{volume}{1}, \bibinfo{pages}{0--4}.
\newblock \DOIprefix\doi{10.7937/K9/TCIA.2016.HJJHBOXZ}.
\bibitem[{Kur{\c{c}} et~al.(2019)Kur{\c{c}}, Sharma, Gupta, Hou, Le, Abousamra,
  Bremer, Birmingham, Diprima, Li, Wang, Balsamo, Bremer, Samaras and
  Saltz}]{TahsinSharma2019}
\bibinfo{author}{Kur{\c{c}}, T.M.}, \bibinfo{author}{Sharma, A.},
  \bibinfo{author}{Gupta, R.}, \bibinfo{author}{Hou, L.}, \bibinfo{author}{Le,
  H.}, \bibinfo{author}{Abousamra, S.}, \bibinfo{author}{Bremer, E.},
  \bibinfo{author}{Birmingham, R.}, \bibinfo{author}{Diprima, T.},
  \bibinfo{author}{Li, N.}, \bibinfo{author}{Wang, F.},
  \bibinfo{author}{Balsamo, J.}, \bibinfo{author}{Bremer, W.},
  \bibinfo{author}{Samaras, D.}, \bibinfo{author}{Saltz, J.H.},
  \bibinfo{year}{2019}.
\newblock \bibinfo{title}{From whole slide tissues to knowledge: Mapping
  sub-cellular morphology of cancer}, in: \bibinfo{booktitle}{Medical Image
  Computing and Computer Assisted Intervention (MICCAI) Brain Lesion (BrainLes)
  workshop}, pp. \bibinfo{pages}{371--379}.
\newblock \DOIprefix\doi{10.1007/978-3-030-46643-5\_37}.
\bibitem[{Liu et~al.(2019)Liu, Xu, Zheng, Gong, Garibaldi, Soria, Green, Ellis,
  Zou and Qiu}]{LiuXu2019}
\bibinfo{author}{Liu, J.}, \bibinfo{author}{Xu, B.}, \bibinfo{author}{Zheng,
  C.}, \bibinfo{author}{Gong, Y.}, \bibinfo{author}{Garibaldi, J.},
  \bibinfo{author}{Soria, D.}, \bibinfo{author}{Green, A.R.},
  \bibinfo{author}{Ellis, I.O.}, \bibinfo{author}{Zou, W.},
  \bibinfo{author}{Qiu, G.}, \bibinfo{year}{2019}.
\newblock \bibinfo{title}{An end-to-end deep learning histochemical scoring
  system for breast cancer {TMA}}.
\newblock \bibinfo{journal}{IEEE Trans Med Imaging} \bibinfo{volume}{38(2)},
  \bibinfo{pages}{617--628}.
\newblock \DOIprefix\doi{10.1109/TMI.2018.2868333}.
\bibitem[{Liu et~al.(2020)Liu, Zhang, Hou, Wang, Mian, Zhang and
  Tang}]{LiuZhang2020}
\bibinfo{author}{Liu, X.}, \bibinfo{author}{Zhang, F.}, \bibinfo{author}{Hou,
  Z.}, \bibinfo{author}{Wang, Z.}, \bibinfo{author}{Mian, L.},
  \bibinfo{author}{Zhang, J.}, \bibinfo{author}{Tang, J.},
  \bibinfo{year}{2020}.
\newblock \bibinfo{title}{Self-supervised learning: Generative or contrastive},
  in: \bibinfo{booktitle}{arXiv}.
\newblock \DOIprefix\doi{10.48550/arXiv.2006.08218}.
\bibitem[{Liu et~al.(2021)Liu, Lin, Cao, Hu, Wei, Zhang, Lin and
  Guo}]{LiuLin2021}
\bibinfo{author}{Liu, Z.}, \bibinfo{author}{Lin, Y.}, \bibinfo{author}{Cao,
  Y.}, \bibinfo{author}{Hu, H.}, \bibinfo{author}{Wei, Y.},
  \bibinfo{author}{Zhang, Z.}, \bibinfo{author}{Lin, S.}, \bibinfo{author}{Guo,
  B.}, \bibinfo{year}{2021}.
\newblock \bibinfo{title}{Swin transformer: Hierarchical vision transformer
  using shifted windows}, in: \bibinfo{booktitle}{IEEE International Conference
  on Computer Vision(ICCV)}, pp. \bibinfo{pages}{9992--10002}.
\newblock \DOIprefix\doi{10.1109/ICCV48922.2021.00986}.
\bibitem[{Lomacenkova and Arandjelovic(2021)}]{LomacenkovaArandjelovic2021}
\bibinfo{author}{Lomacenkova, A.}, \bibinfo{author}{Arandjelovic, O.},
  \bibinfo{year}{2021}.
\newblock \bibinfo{title}{Whole slide pathology image patch based deep
  classification: An investigation of the effects of the latent autoencoder
  representation and the loss function form}, in: \bibinfo{booktitle}{IEEE
  Biomedical and Health Informatics (BHI)}, pp. \bibinfo{pages}{1--4}.
\newblock \DOIprefix\doi{10.1109/BHI50953.2021.9508577}.
\bibitem[{Long et~al.(2015)Long, Shelhamer and Darrell}]{LongShelhamer2015}
\bibinfo{author}{Long, J.}, \bibinfo{author}{Shelhamer, E.},
  \bibinfo{author}{Darrell, T.}, \bibinfo{year}{2015}.
\newblock \bibinfo{title}{Fully convolutional networks for semantic
  segmentation}, in: \bibinfo{booktitle}{IEEE Conference on Computer Vision and
  Pattern Recognition (CVPR)}, \bibinfo{publisher}{IEEE Computer Society}. pp.
  \bibinfo{pages}{3431--3440}.
\newblock \DOIprefix\doi{10.1109/CVPR.2015.7298965}.
\bibitem[{Macenko et~al.(2009)Macenko, Niethammer, Marron, Borland, Woosley,
  Guan, Schmitt and Thomas}]{Macenko}
\bibinfo{author}{Macenko, M.}, \bibinfo{author}{Niethammer, M.},
  \bibinfo{author}{Marron, J.S.}, \bibinfo{author}{Borland, D.},
  \bibinfo{author}{Woosley, J.T.}, \bibinfo{author}{Guan, X.},
  \bibinfo{author}{Schmitt, C.}, \bibinfo{author}{Thomas, N.E.},
  \bibinfo{year}{2009}.
\newblock \bibinfo{title}{A method for normalizing histology slides for
  quantitative analysis}, in: \bibinfo{booktitle}{IEEE International Symposium
  on Biomedical Imaging (ISBI)}, pp. \bibinfo{pages}{1107--1110}.
\newblock \DOIprefix\doi{10.1109/ISBI.2009.5193250}.
\bibitem[{Milletari et~al.(2016)Milletari, Navab and
  Ahmadi}]{MilletariNavab2016}
\bibinfo{author}{Milletari, F.}, \bibinfo{author}{Navab, N.},
  \bibinfo{author}{Ahmadi, S.}, \bibinfo{year}{2016}.
\newblock \bibinfo{title}{V-net: Fully convolutional neural networks for
  volumetric medical image segmentation}, in: \bibinfo{booktitle}{3D Vision
  (3DV)}, pp. \bibinfo{pages}{565--571}.
\newblock \DOIprefix\doi{10.1109/3DV.2016.79}.
\bibitem[{Mohammad et~al.(2017)Mohammad, Peikari, Sherine, Salama, Sharon,
  Nofech-Mozes, Anne, L. and Martel}]{MohammadPeikari2017}
\bibinfo{author}{Mohammad}, \bibinfo{author}{Peikari},
  \bibinfo{author}{Sherine}, \bibinfo{author}{Salama},
  \bibinfo{author}{Sharon}, \bibinfo{author}{Nofech-Mozes},
  \bibinfo{author}{Anne}, \bibinfo{author}{L.}, \bibinfo{author}{Martel},
  \bibinfo{year}{2017}.
\newblock \bibinfo{title}{Automatic cellularity assessment from post-treated
  breast surgical specimens}.
\newblock \bibinfo{journal}{Cytometry Part A} \bibinfo{volume}{4(1)}.
\newblock \DOIprefix\doi{10.1002/cyto.a.23244}.
\bibitem[{Nadeem et~al.(2020)Nadeem, Hollmann and Tannenbaum}]{Saad}
\bibinfo{author}{Nadeem, S.}, \bibinfo{author}{Hollmann, T.},
  \bibinfo{author}{Tannenbaum, A.}, \bibinfo{year}{2020}.
\newblock \bibinfo{title}{Multimarginal wasserstein barycenter for stain
  normalization and augmentation}, in: \bibinfo{booktitle}{Medical Image
  Computing and Computer-Assisted Intervention (MICCAI)}, pp.
  \bibinfo{pages}{6860--6868}.
\newblock \DOIprefix\doi{10.1007/978-3-030-59722-1_35}.
\bibitem[{Naylor et~al.(2019a)Naylor, Lae, Reyal and Walter}]{NaylorMarick2019}
\bibinfo{author}{Naylor, P.}, \bibinfo{author}{Lae, M.},
  \bibinfo{author}{Reyal, F.}, \bibinfo{author}{Walter, T.},
  \bibinfo{year}{2019}a.
\newblock \bibinfo{title}{Segmentation of nuclei in histopathology images by
  deep regression of the distance map}.
\newblock \bibinfo{journal}{IEEE Trans Med Imaging} \bibinfo{volume}{38(2)},
  \bibinfo{pages}{448--459}.
\newblock \DOIprefix\doi{10.1109/TMI.2018.2865709}.
\bibitem[{Naylor et~al.(2019b)Naylor, Lae, Reyal and Walter}]{NaylorLae2019}
\bibinfo{author}{Naylor, P.}, \bibinfo{author}{Lae, M.},
  \bibinfo{author}{Reyal, F.}, \bibinfo{author}{Walter, T.},
  \bibinfo{year}{2019}b.
\newblock \bibinfo{title}{Segmentation of nuclei in histopathology images by
  deep regression of the distance map}.
\newblock \bibinfo{journal}{IEEE Trans Med Imaging} \bibinfo{volume}{38(2)},
  \bibinfo{pages}{448--459}.
\newblock \DOIprefix\doi{10.1109/TMI.2018.2865709}.
\bibitem[{Nicolas et~al.(2018)Nicolas, Santiago, Theodore, Navneet, Matija,
  David, Moreira, Narges and Aristotelis}]{NicolasSantiago2018}
\bibinfo{author}{Nicolas, C.}, \bibinfo{author}{Santiago, O.P.},
  \bibinfo{author}{Theodore, S.}, \bibinfo{author}{Navneet, N.},
  \bibinfo{author}{Matija, S.}, \bibinfo{author}{David, F.},
  \bibinfo{author}{Moreira, A.L.}, \bibinfo{author}{Narges, R.},
  \bibinfo{author}{Aristotelis, T.}, \bibinfo{year}{2018}.
\newblock \bibinfo{title}{Classification and mutation prediction from
  non–small cell lung cancer histopathology images using deep learning}.
\newblock \bibinfo{journal}{Nature Medicine} \bibinfo{volume}{24(5)},
  \bibinfo{pages}{1559--1567}.
\newblock \DOIprefix\doi{10.1038/s41591-018-0177-5}.
\bibitem[{Nishar et~al.(2020)Nishar, Chavanke and Singhal}]{Harshal}
\bibinfo{author}{Nishar, H.}, \bibinfo{author}{Chavanke, N.},
  \bibinfo{author}{Singhal, N.}, \bibinfo{year}{2020}.
\newblock \bibinfo{title}{Histopathological stain transfer using style transfer
  network with adversarial loss}, in: \bibinfo{booktitle}{Medical Image
  Computing and Computer-Assisted Intervention (MICCAI)}, pp.
  \bibinfo{pages}{6860--6868}.
\newblock \DOIprefix\doi{10.1007/978-3-030-59722-1_32}.
\bibitem[{Rabinovich et~al.(2003)Rabinovich, Agarwal, Laris, Price and
  Belongie}]{RabinovichAgarwal2003}
\bibinfo{author}{Rabinovich, A.}, \bibinfo{author}{Agarwal, S.},
  \bibinfo{author}{Laris, C.}, \bibinfo{author}{Price, J.H.},
  \bibinfo{author}{Belongie, S.J.}, \bibinfo{year}{2003}.
\newblock \bibinfo{title}{Unsupervised color decomposition of histologically
  stained tissue samples}, in: \bibinfo{editor}{Thrun, S.},
  \bibinfo{editor}{Saul, L.K.}, \bibinfo{editor}{Sch{\"{o}}lkopf, B.} (Eds.),
  \bibinfo{booktitle}{Neural Information Processing Systems (NIPS)}, pp.
  \bibinfo{pages}{667--674}.
\newblock \DOIprefix\doi{10.5555/2981345.2981429}.
\bibitem[{Radford et~al.(2016)Radford, Metz and Chintala}]{RadfordMetz2016}
\bibinfo{author}{Radford, A.}, \bibinfo{author}{Metz, L.},
  \bibinfo{author}{Chintala, S.}, \bibinfo{year}{2016}.
\newblock \bibinfo{title}{Unsupervised representation learning with deep
  convolutional generative adversarial networks}, in: \bibinfo{editor}{Bengio,
  Y.}, \bibinfo{editor}{LeCun, Y.} (Eds.), \bibinfo{booktitle}{International
  Conference on Learning Representations (ICLR)}.
\newblock \DOIprefix\doi{10.48550/arXiv.1511.06434}.
\bibitem[{Reinhard et~al.(2001)Reinhard, Adhikhmin, Gooch and
  Shirley}]{Reinhard}
\bibinfo{author}{Reinhard, E.}, \bibinfo{author}{Adhikhmin, M.},
  \bibinfo{author}{Gooch, B.}, \bibinfo{author}{Shirley, P.},
  \bibinfo{year}{2001}.
\newblock \bibinfo{title}{Color transfer between images}.
\newblock \bibinfo{journal}{IEEE Computer Graphics and Applications}
  \bibinfo{volume}{21(5)}, \bibinfo{pages}{34--41}.
\newblock \DOIprefix\doi{10.1109/38.946629}.
\bibitem[{Ronneberger et~al.(2015)Ronneberger, Fischer and
  Brox}]{RonnebergerFischer2015}
\bibinfo{author}{Ronneberger, O.}, \bibinfo{author}{Fischer, P.},
  \bibinfo{author}{Brox, T.}, \bibinfo{year}{2015}.
\newblock \bibinfo{title}{U-net: Convolutional networks for biomedical image
  segmentation}, in: \bibinfo{booktitle}{Medical Image Computing and
  Computer-Assisted Intervention (MICCAI)}, pp. \bibinfo{pages}{234--241}.
\newblock \DOIprefix\doi{10.1007/978-3-319-24574-4\_28}.
\bibitem[{Schmitz et~al.(2021)Schmitz, Madesta, Nielsen, Krause, Steurer,
  Werner and R{\"{o}}sch}]{SchmitzMadesta2021}
\bibinfo{author}{Schmitz, R.}, \bibinfo{author}{Madesta, F.},
  \bibinfo{author}{Nielsen, M.}, \bibinfo{author}{Krause, J.},
  \bibinfo{author}{Steurer, S.}, \bibinfo{author}{Werner, R.},
  \bibinfo{author}{R{\"{o}}sch, T.}, \bibinfo{year}{2021}.
\newblock \bibinfo{title}{Multi-scale fully convolutional neural networks for
  histopathology image segmentation: From nuclear aberrations to the global
  tissue architecture}.
\newblock \bibinfo{journal}{Med Image Anal} \bibinfo{volume}{70},
  \bibinfo{pages}{101996}.
\newblock \DOIprefix\doi{10.1016/j.media.2021.101996}.
\bibitem[{Shaban et~al.(2019)Shaban, Baur, Navab and Albarqouni}]{Staingan}
\bibinfo{author}{Shaban, M.T.}, \bibinfo{author}{Baur, C.},
  \bibinfo{author}{Navab, N.}, \bibinfo{author}{Albarqouni, S.},
  \bibinfo{year}{2019}.
\newblock \bibinfo{title}{Staingan: Stain style transfer for digital
  histological images}, in: \bibinfo{booktitle}{IEEE International Symposium on
  Biomedical Imaging (ISBI)}, pp. \bibinfo{pages}{953--956}.
\newblock \DOIprefix\doi{10.1109/ISBI.2019.8759152}.
\bibitem[{Song et~al.()Song, Tan, Jiang, Cheng, Lei and Wang}]{SongTan2019}
\bibinfo{author}{Song, Y.}, \bibinfo{author}{Tan, E.}, \bibinfo{author}{Jiang,
  X.}, \bibinfo{author}{Cheng, J.}, \bibinfo{author}{Lei, B.},
  \bibinfo{author}{Wang, T.}, .
\newblock \bibinfo{title}{Corrections to "accurate cervical cell segmentation
  from overlapping clumps in pap smear images"}.
\newblock \bibinfo{journal}{IEEE Trans Med Imaging} \bibinfo{volume}{38(6)},
  \bibinfo{pages}{1543--1558}.
\newblock \DOIprefix\doi{10.1109/TMI.2019.2913056}.
\bibitem[{Symmans et~al.(2007)Symmans, Peintinger, Hatzis, Rajan, Kuerer,
  Valero, Assad, Poniecka, Hennessy and Green}]{SymmansPeintinger2007}
\bibinfo{author}{Symmans, W.F.}, \bibinfo{author}{Peintinger, F.},
  \bibinfo{author}{Hatzis, C.}, \bibinfo{author}{Rajan, R.},
  \bibinfo{author}{Kuerer, H.}, \bibinfo{author}{Valero, V.},
  \bibinfo{author}{Assad, L.}, \bibinfo{author}{Poniecka, A.},
  \bibinfo{author}{Hennessy, B.}, \bibinfo{author}{Green, M.},
  \bibinfo{year}{2007}.
\newblock \bibinfo{title}{Measurement of residual breast cancer burden to
  predict survival after neoadjuvant chemotherapy.}
\newblock \bibinfo{journal}{Journal of Clinical Oncology}
  \bibinfo{volume}{25(28)}, \bibinfo{pages}{4414--4422}.
\bibitem[{Tan and Le(2019)}]{TanLe2019}
\bibinfo{author}{Tan, M.}, \bibinfo{author}{Le, Q.}, \bibinfo{year}{2019}.
\newblock \bibinfo{title}{Efficientnet: Rethinking model scaling for
  convolutional neural networks}, in: \bibinfo{editor}{Chaudhuri, K.},
  \bibinfo{editor}{Salakhutdinov, R.} (Eds.), \bibinfo{booktitle}{International
  Conference on Machine Learning (ICML)}, \bibinfo{publisher}{PMLR}. pp.
  \bibinfo{pages}{6105--6114}.
\newblock \URLprefix \url{https://proceedings.mlr.press/v97/tan19a.html}.
\bibitem[{Tellez et~al.(2018)Tellez, Balkenhol, Otte{-}H{\"{o}}ller, van~de
  Loo, Vogels, Bult, Wauters, Vreuls, Mol, Karssemeijer, Litjens, van~der Laak
  and Ciompi}]{TellezBalkenhol2018}
\bibinfo{author}{Tellez, D.}, \bibinfo{author}{Balkenhol, M.},
  \bibinfo{author}{Otte{-}H{\"{o}}ller, I.}, \bibinfo{author}{van~de Loo, R.},
  \bibinfo{author}{Vogels, R.}, \bibinfo{author}{Bult, P.},
  \bibinfo{author}{Wauters, C.}, \bibinfo{author}{Vreuls, W.},
  \bibinfo{author}{Mol, S.}, \bibinfo{author}{Karssemeijer, N.},
  \bibinfo{author}{Litjens, G.}, \bibinfo{author}{van~der Laak, J.},
  \bibinfo{author}{Ciompi, F.}, \bibinfo{year}{2018}.
\newblock \bibinfo{title}{Whole-slide mitosis detection in h{\&}e breast
  histology using {PHH3} as a reference to train distilled stain-invariant
  convolutional networks}.
\newblock \bibinfo{journal}{IEEE Trans Med Imaging} \bibinfo{volume}{37(9)},
  \bibinfo{pages}{2126--2136}.
\newblock \DOIprefix\doi{10.1109/TMI.2018.2820199}.
\bibitem[{Vahadane et~al.(2016)Vahadane, Peng, Sethi, Albarqouni, Wang, Baust,
  Steiger, Schlitter, Esposito and Navab}]{Vahadane}
\bibinfo{author}{Vahadane, A.}, \bibinfo{author}{Peng, T.},
  \bibinfo{author}{Sethi, A.}, \bibinfo{author}{Albarqouni, S.},
  \bibinfo{author}{Wang, L.}, \bibinfo{author}{Baust, M.},
  \bibinfo{author}{Steiger, K.}, \bibinfo{author}{Schlitter, A.M.},
  \bibinfo{author}{Esposito, I.}, \bibinfo{author}{Navab, N.},
  \bibinfo{year}{2016}.
\newblock \bibinfo{title}{Structure-preserving color normalization and sparse
  stain separation for histological images}.
\newblock \bibinfo{journal}{IEEE Trans Med Imaging} \bibinfo{volume}{35},
  \bibinfo{pages}{1962--1971}.
\newblock \DOIprefix\doi{10.1109/TMI.2016.2529665}.
\bibitem[{Wang et~al.(2014)Wang, Huang, Wang and Wang}]{WangHuang2014}
\bibinfo{author}{Wang, W.}, \bibinfo{author}{Huang, Y.}, \bibinfo{author}{Wang,
  Y.}, \bibinfo{author}{Wang, L.}, \bibinfo{year}{2014}.
\newblock \bibinfo{title}{Generalized autoencoder: {A} neural network framework
  for dimensionality reduction}, in: \bibinfo{booktitle}{IEEE Computer Vision
  and Pattern Recognition (CVPR)}, pp. \bibinfo{pages}{496--503}.
\newblock \DOIprefix\doi{10.1109/CVPRW.2014.79}.
\bibitem[{Wang et~al.(2020)Wang, Chen, Gan, Lin, Dou, Tsougenis, Huang, Cai and
  Heng}]{WangChen2020}
\bibinfo{author}{Wang, X.}, \bibinfo{author}{Chen, H.}, \bibinfo{author}{Gan,
  C.}, \bibinfo{author}{Lin, H.}, \bibinfo{author}{Dou, Q.},
  \bibinfo{author}{Tsougenis, E.}, \bibinfo{author}{Huang, Q.},
  \bibinfo{author}{Cai, M.}, \bibinfo{author}{Heng, P.}, \bibinfo{year}{2020}.
\newblock \bibinfo{title}{Weakly supervised deep learning for whole slide lung
  cancer image analysis}.
\newblock \bibinfo{journal}{IEEE Trans Cybern} \bibinfo{volume}{50(9)},
  \bibinfo{pages}{3950--3962}.
\newblock \DOIprefix\doi{10.1109/TCYB.2019.2935141}.
\bibitem[{Woo et~al.(2018)Woo, Park, Lee and Kweon}]{WooPark2018}
\bibinfo{author}{Woo, S.}, \bibinfo{author}{Park, J.}, \bibinfo{author}{Lee,
  J.}, \bibinfo{author}{Kweon, I.S.}, \bibinfo{year}{2018}.
\newblock \bibinfo{title}{{CBAM:} convolutional block attention module}, in:
  \bibinfo{booktitle}{European Conference Computer Vision (ECCV)}, pp.
  \bibinfo{pages}{3--19}.
\newblock \DOIprefix\doi{10.1007/978-3-030-01234-2\_1}.
\bibitem[{Xie and Tu(2017)}]{XieTu2017}
\bibinfo{author}{Xie, S.}, \bibinfo{author}{Tu, Z.}, \bibinfo{year}{2017}.
\newblock \bibinfo{title}{Holistically-nested edge detection}.
\newblock \bibinfo{journal}{Int. J. Comput. Vis.} \bibinfo{volume}{125(3)},
  \bibinfo{pages}{3--18}.
\newblock \DOIprefix\doi{10.1007/s11263-017-1004-z}.
\bibitem[{Xie et~al.(2015)Xie, Kong, Xing, Liu, Su and Yang}]{XieKong2015}
\bibinfo{author}{Xie, Y.}, \bibinfo{author}{Kong, X.}, \bibinfo{author}{Xing,
  F.}, \bibinfo{author}{Liu, F.}, \bibinfo{author}{Su, H.},
  \bibinfo{author}{Yang, L.}, \bibinfo{year}{2015}.
\newblock \bibinfo{title}{Deep voting: {A} robust approach toward nucleus
  localization in microscopy images}, in: \bibinfo{booktitle}{Medical Image
  Computing and Computer-Assisted Intervention (MICCAI)}, pp.
  \bibinfo{pages}{374--382}.
\newblock \DOIprefix\doi{10.1007/978-3-319-24574-4\_45}.
\bibitem[{Xu et~al.(2019)Xu, Liu, Hou, Liu, Garibaldi, Ellis, Green, Shen and
  Qiu}]{XuLiu2019}
\bibinfo{author}{Xu, B.}, \bibinfo{author}{Liu, J.}, \bibinfo{author}{Hou, X.},
  \bibinfo{author}{Liu, B.}, \bibinfo{author}{Garibaldi, J.},
  \bibinfo{author}{Ellis, I.O.}, \bibinfo{author}{Green, A.},
  \bibinfo{author}{Shen, L.}, \bibinfo{author}{Qiu, G.}, \bibinfo{year}{2019}.
\newblock \bibinfo{title}{Look, investigate, and classify: {A} deep hybrid
  attention method for breast cancer classification}, in:
  \bibinfo{booktitle}{IEEE International Symposium on Biomedical Imaging
  (ISBI)}, pp. \bibinfo{pages}{914--918}.
\newblock \DOIprefix\doi{10.1109/ISBI.2019.8759454}.
\bibitem[{Xu et~al.(2020)Xu, Gopale, Zhang, Brown, Begoli and
  Bethard}]{XuGopale2020}
\bibinfo{author}{Xu, D.}, \bibinfo{author}{Gopale, M.}, \bibinfo{author}{Zhang,
  J.}, \bibinfo{author}{Brown, K.}, \bibinfo{author}{Begoli, E.},
  \bibinfo{author}{Bethard, S.}, \bibinfo{year}{2020}.
\newblock \bibinfo{title}{Unified medical language system resources improve
  sieve-based generation and bidirectional encoder representations from
  transformers (bert)-based ranking for concept normalization}.
\newblock \bibinfo{journal}{J. Am. Medical Informatics Assoc.}
  \bibinfo{volume}{27(10)}, \bibinfo{pages}{1510--1519}.
\newblock \DOIprefix\doi{10.1093/jamia/ocaa080}.
\bibitem[{Zhang et~al.(2016)Zhang, Isola and Efros}]{ZhangIsola2016}
\bibinfo{author}{Zhang, R.}, \bibinfo{author}{Isola, P.},
  \bibinfo{author}{Efros, A.A.}, \bibinfo{year}{2016}.
\newblock \bibinfo{title}{Colorful image colorization}, in:
  \bibinfo{editor}{Leibe, B.}, \bibinfo{editor}{Matas, J.},
  \bibinfo{editor}{Sebe, N.}, \bibinfo{editor}{Welling, M.} (Eds.),
  \bibinfo{booktitle}{European Conference Computer Vision (ECCV)}, pp.
  \bibinfo{pages}{649--666}.
\newblock \DOIprefix\doi{10.1007/978-3-319-46487-9\_40}.
\bibitem[{Zhang et~al.(2017)Zhang, Isola and Efros}]{ZhangIsola2017}
\bibinfo{author}{Zhang, R.}, \bibinfo{author}{Isola, P.},
  \bibinfo{author}{Efros, A.A.}, \bibinfo{year}{2017}.
\newblock \bibinfo{title}{Split-brain autoencoders: Unsupervised learning by
  cross-channel prediction}, in: \bibinfo{booktitle}{IEEE Computer Vision and
  Pattern Recognition (CVPR)}, pp. \bibinfo{pages}{645--654}.
\newblock \DOIprefix\doi{10.1109/CVPR.2017.76}.
\bibitem[{Zheng et~al.(2021)Zheng, Lu, Zhao, Zhu, Luo, Wang, Fu, Feng, Xiang,
  Torr and Zhang}]{ZhengLu2021}
\bibinfo{author}{Zheng, S.}, \bibinfo{author}{Lu, J.}, \bibinfo{author}{Zhao,
  H.}, \bibinfo{author}{Zhu, X.}, \bibinfo{author}{Luo, Z.},
  \bibinfo{author}{Wang, Y.}, \bibinfo{author}{Fu, Y.}, \bibinfo{author}{Feng,
  J.}, \bibinfo{author}{Xiang, T.}, \bibinfo{author}{Torr, P.H.S.},
  \bibinfo{author}{Zhang, L.}, \bibinfo{year}{2021}.
\newblock \bibinfo{title}{Rethinking semantic segmentation from a
  sequence-to-sequence perspective with transformers}, in:
  \bibinfo{booktitle}{IEEE Computer Vision and Pattern Recognition (CVPR)}, pp.
  \bibinfo{pages}{6881--6890}.
\newblock \DOIprefix\doi{10.48550/arXiv.2012.15840}.
\bibitem[{Zhou et~al.(2016)Zhou, Khosla, Lapedriza, Oliva and
  Torralba}]{ZhouKhosla2016}
\bibinfo{author}{Zhou, B.}, \bibinfo{author}{Khosla, A.},
  \bibinfo{author}{Lapedriza, {\`{A}}.}, \bibinfo{author}{Oliva, A.},
  \bibinfo{author}{Torralba, A.}, \bibinfo{year}{2016}.
\newblock \bibinfo{title}{Learning deep features for discriminative
  localization}, in: \bibinfo{booktitle}{IEEE Computer Vision and Pattern
  Recognition (CVPR)}, pp. \bibinfo{pages}{2921--2929}.
\newblock \DOIprefix\doi{10.1109/CVPR.2016.319}.
\bibitem[{Zhou et~al.(2018)Zhou, Siddiquee, Tajbakhsh and
  Liang}]{ZhouSiddiquee2018}
\bibinfo{author}{Zhou, Z.}, \bibinfo{author}{Siddiquee, M.M.R.},
  \bibinfo{author}{Tajbakhsh, N.}, \bibinfo{author}{Liang, J.},
  \bibinfo{year}{2018}.
\newblock \bibinfo{title}{Unet++: {A} nested u-net architecture for medical
  image segmentation}, in: \bibinfo{booktitle}{Medical Image Computing and
  Computer Assisted Intervention (MICCAI) DLMIA workshop}, pp.
  \bibinfo{pages}{3--11}.
\newblock \DOIprefix\doi{10.1007/978-3-030-00889-5\_1}.
\bibitem[{Zoph et~al.(2018)Zoph, Vasudevan, Shlens and Le}]{ZophVasudevan2018}
\bibinfo{author}{Zoph, B.}, \bibinfo{author}{Vasudevan, V.},
  \bibinfo{author}{Shlens, J.}, \bibinfo{author}{Le, Q.V.},
  \bibinfo{year}{2018}.
\newblock \bibinfo{title}{Learning transferable architectures for scalable
  image recognition}, in: \bibinfo{booktitle}{IEEE Conference on Computer
  Vision and Pattern Recognition (CVPR)}, pp. \bibinfo{pages}{8697--8710}.
\newblock \DOIprefix\doi{10.1109/CVPR.2018.00907}.

\end{thebibliography}

\end{document}